\documentclass[iop]{emulateapj}

\accepted{\apj}

\shorttitle{Multi-wavelength observations of the black widow pulsar 2FGL
J2339.6$-$0532} \shortauthors{Yatsu et al.}
\usepackage{natbib}
\usepackage{graphicx}	
\usepackage{url}	
\begin{document}
\title{Multi-wavelength observations of the black widow pulsar 2FGL
J2339.6$-$0532 with OISTER and Suzaku}


\author{
Yoichi YATSU\altaffilmark{1},
Jun KATAOKA\altaffilmark{2},
Yosuke TAKAHASHI\altaffilmark{2},
Yutaro TACHIBANA\altaffilmark{1},
Nobuyuki KAWAI\altaffilmark{1},
Shimpei SHIBATA\altaffilmark{3},
Sean PIKE\altaffilmark{4},
Taketoshi YOSHII\altaffilmark{1},
Makoto ARIMOTO\altaffilmark{1}, 
Yoshihiko SAITO\altaffilmark{1}, 
Takeshi NAKAMORI\altaffilmark{3},
Kazuhiro SEKIGUCHI\altaffilmark{5}, 
Daisuke KURODA\altaffilmark{6},
Kenshi YANAGISAWA\altaffilmark{6},
Hidekazu HANAYAMA\altaffilmark{7}
Makoto WATANABE\altaffilmark{8},
Ko HAMAMOTO\altaffilmark{8},
Hikaru NAKAO\altaffilmark{8},
Akihito OZAKI\altaffilmark{8},
Kentaro MOTOHARA\altaffilmark{9},
Masahiro KONISHI\altaffilmark{9},
Ken TATEUCHI\altaffilmark{9},
Noriyuki MATSUNAGA\altaffilmark{10},
Tomoki MOROKUMA\altaffilmark{9},
Takahiro NAGAYAMA\altaffilmark{11,12}
Katsuhiro MURATA\altaffilmark{11}
Hiroshi AKITAYA\altaffilmark{13},
Michitoshi YOSHIDA\altaffilmark{13},
Gamal B.~ALI\altaffilmark{14},
A.~Essam MOHAMED\altaffilmark{14},
Mizuki ISOGAI\altaffilmark{5,15}
Akira ARAI\altaffilmark{15}
Hidenori TAKAHASHI\altaffilmark{9,16},
Osamu HASHIMOTO\altaffilmark{16},
Ryo MIYANOSHITA\altaffilmark{12}
Toshihiro OMODAKA\altaffilmark{12},
Jun TAKAHASHI\altaffilmark{17},
Noritaka TOKIMASA\altaffilmark{17},
Kentaro MATSUDA\altaffilmark{17},
Shin-ichiro OKUMURA\altaffilmark{18} 
Kota NISHIYAMA\altaffilmark{18} 
Seitaro URAKAWA\altaffilmark{18} 
Daisaku NOGAMI\altaffilmark{19},
Yumiko OASA\altaffilmark{20},
\protect \and
on behalf of OISTER team\altaffilmark{21}
}

\altaffiltext{1}{Department of Physics, Tokyo Institute of Technology,
2-12-1 Ohokayama Meguro Tokyo 152-8551, Japan} 
\email{yatsu@hp.phys.titech.ac.jp}

\altaffiltext{2}{Research Institute for Science and Engineering, Waseda
University, 3-4-1 Ookubo Shinjuku Tokyo 169-8555, Japan}

\altaffiltext{3}{Department of Physics, Yamagata University, 1-4-12
Kojirakawamachi, Yamagata, Yamagata 990-8560, Japan}

\altaffiltext{4}{Department of Physics, Brown University, 182 Hope
Street, Providence, Rhode Island 02912, USA}

\altaffiltext{5}{National Astronomical Observatory of Japan, National
Institute of Natural Sciences, 2-21-1 Osawa, Mitaka, Tokyo 181.8588,
Japan}

\altaffiltext{6}{Okayama Astrophysical Observatory, National
Astronomical Observatory of Japan, 3037-5 Honjo, Kamogata, Asakuchi,
Okayama 719-0232}

\altaffiltext{7}{Ishigakijima Astronomical Observatory, National
Astronomical Observatory of Japan, Ishigaki, Okinawa 907-0024, Japan}

\altaffiltext{8}{Department of Cosmosciences, Hokkaido University, Kita
10, Nishi 8, Kita-ku, Sapporo, Hokkaido 060-0810, Japan}

\altaffiltext{9}{Institute of Astronomy, Graduate School of Science, The
University of Tokyo, 2-21-1 Osawa, Mitaka, Tokyo 181-0015, Japan}

\altaffiltext{10}{Department of Astronomy, Graduate School of Science, The
University of Tokyo, 7-3-1 Hongo, Bunkyo-ku, Tokyo 113-0033, Japan}

\altaffiltext{11}{Department of Astrophysics, Nagoya University,
Furo-cho, Chikusa-ku, Nagoya 464-8602, Japan}

\altaffiltext{12}{Faculty of Science, Kagoshima University, 1-21-30
Korimoto, Kagoshima, Kagoshima 890-0065, Japan}

\altaffiltext{13}{Hiroshima Astrophysical Science Center, Hiroshima University,
1-3-1 Kagamiyama, Higashihiroshima, Hiroshima 739-8526, Japan}

\altaffiltext{14}{National Research Institute of Astronomy and Geophysics, Cairo
11722, Egypt}

\altaffiltext{15}{Kamiyama Astronomical Observatory, Kyoto Sangyo
University, Kamigamomotoyama, Kita-ku, Kyoto, Kyoto 603-8555, Japan}

\altaffiltext{16}{Gunma Astronomical Observatory,
6860-86 Nakayama, Takayama-mura, Agatsuma-gun Gunma 377-0702, Japan}

\altaffiltext{17}{Nishi-Harima Astronomical Observatory, Center for
Astronomy, University of Hyogo, 407-2, Nishigaichi, Sayo-cho, Sayo,
Hyogo 679-5313, Japan}

\altaffiltext{18}{Bisei Spaceguard Center, Japan Spaceguard Association,
1716-3 Okura, Bisei-cho, Ibara, Okayama 714-1411, Japan}

\altaffiltext{19}{Department of Astronomy, Kyoto University, Kitashirakawa Oiwake-cho, Sakyo-ku, Kyoto, Kyoto 606-8502, Japan}

\altaffiltext{20}{Faculty of Education, Saitama University, 255
Shimo-Okubo, Sakura, Saitama, Saitama 388-8570, Japan}

\altaffiltext{21}{Optical and Infrared Synergetic Telescopes for
Education and Research}




\begin{abstract}
Multi-wavelength observations of the black-widow binary system 2FGL
J2339.6$-$0532 are reported.  The Fermi gamma-ray source 2FGL
J2339.6-0532 was recently categorized as a black widow in which a
recycled millisecond pulsar (MSP) is evaporating up the companion star
with its powerful pulsar wind.  Our optical observations show clear
sinusoidal light curves due to the asymmetric temperature distribution
of the companion star.  Assuming a simple geometry, we constrained the
range of the inclination angle of the binary system to $52\degr < i <
59\degr$, which enables us to discuss the interaction between the pulsar
wind and the companion in detail. The X-ray spectrum consists of two
components: a soft, steady component that seems to originate from the
surface of the MSP, and a hard variable component from the
wind-termination shock near the companion star.  The measured X-ray
luminosity is comparable to the bolometric luminosity of the companion,
meaning that the heating efficiency is less than 0.5.  In the companion
orbit, $10^{11}$ cm from the pulsar, the pulsar wind is already in
particle dominant-stage, with a magnetization parameter of $\sigma <
0.1$.
In addition, we precisely investigated the time variations of the X-ray
periodograms and detected a weakening of orbital modulation.  The
observed phenomenon may be related to an unstable pulsar-wind activity
or a weak mass accretion, both of which can result in the temporal
extinction of radio-pulse.
\end{abstract}





\section{Introduction}
A milli-second pulsar (MSP) is believed to evolve from a dead pulsar in
a binary system, via mass accretion from a companion star that has spun
up the pulsar for billions of years \citep{1982Natur.300..728A}.  Indeed
most MSPs are discovered in binary systems, however isolated MSPs also
exist.  The missing link between the isolated MSPs and binary MSPs
is black widow pulsars in which a sufficiently spun up pulsar is
evaporating its companion star with its powerful pulsar wind.  The
prototype of the black widow is PSR B1957$+$20 with a 1.61 ms radio
pulse \citep{1988Natur.333..237F}.  So far less than 10 black widow
class objects have been discovered however the evolution process from an
accreting MSP to a rotation-powered MSP is still
unclear\citep{2011AIPC.1357..127R,2013IAUS..291..127R}.

Moreover a black widow may be a good probe for deep investigation of
pulsar wind.  Currently young rotation powered pulsars are believed to
have pulsar wind nebulae (PWNe) made up of relativistic
electron-positron plasma, as seen in the Crab nebula
\citep{2000ApJ...536L..81W, 2002ApJ...577L..49H}, however the physical
mechanism that generates the pulsar wind is not yet understood.
Although we do not know whether the descriptions of the Crab pulsar can
be applied to a recycled MSP with magnetic field four orders of
magnitude weaker than that of typical radio pulsars, if this is the
case, the black widows can provide a crucial chance for us to probe
pulsar wind via the direct interaction with the companion nearby the
light cylinder.  Theoretically pulsar wind is dominated by Poynting flux
at the light cylinder, therefore the magnetic energy in the wind must be
converted into kinetic energy just after the wind flies out from the
light cylinder as reported by \cite{2012Natur.482..507A}, and we can
approach the origin of the pulsar wind much deeper with black widows for
investigating the unknown physical mechanism which can explain the
$\sigma$ paradox.  In order to clarify the history of MSP formation and
also to constrain the physical mechanism of particle acceleration
“just” around the pulsar, a black widow pulsar is an intriguing
target.

The Large Area Telescope (LAT) on the {\itshape Fermi} satellite, with
unprecedented sensitivity and angular resolution in the energy range of
100 MeV to greater than 300 GeV, has discovered more than 2000 gamma-ray
sources since its launch, and 30\% of these sources are still
unidentified.  A bright gamma-ray source discovered at high galactic
latitude vicinity, 2FGL J2339.6$-$0532 (1FGL J 2339.7$-$0531), was also
listed in the first Fermi source catalog as an un-identified source
\citep{2010ApJS..188..405A, 2012ApJ...753...83A}.  The gamma-ray flux
amounts to $3.0\pm0.2\times10^{-11}$ erg s$^{-1}$ cm$^{-2}$ with a
variable index of 15.7, indicating that the gamma-ray flux seems to be
steady at the month time scale.  While the gamma-ray spectrum has a
cutoff structure at 3 GeV.  These characteristics lead us to believe the
object is a pulsar.  A follow-up X-ray observation conducted with
{\itshape Chandra} discovered an X-ray point source within the error
circle expected from the gamma-ray image\citep{2012ApJ...747L...3K}.
However a radio pulse was not detected at the position at that moment.
On the other hand, ground based optical observations discovered clear
sinusoidal variability with a period of 4.63 hours, which implies the
object is in a binary system and the observed optical variability is
likely related to the orbital motion.  The intensity of the optical
counterpart varies from 20 to 17 mag in
R-band\citep{2011ApJ...743L..26R,2012ApJ...747L...3K}.  Moreover the
phase-resolved spectroscopy indicated that the companion may be K-class
star with a mass of 0.075 $M_{\sun}$.  This means that the pulsar side
hemisphere of the companion star is drastically heated and the energy
might be supplied from an unknown recycled MSP via pulsar wind, as seen
in PSR B1957$+$20.

In this paper, we report multi-wavelength observations of the newly
discovered black widow binary system 2FGL J2339.6-0532, covering
near-infrared, optical and X-ray energy band.  In section 2 and 3, the
optical and X-ray observations and the obtained results are described,
respectively.  In section 4, we discuss the orbital parameters based on
the obtained phase resolved SED and the properties of the pulsar wind
just around the pulsar as well as an interpretation of the X-ray light
curve showing intriguing irregular variability.

\section{Optical observations}
\subsection{Observation}
\begin{figure*}[t]
 \begin{center}
\includegraphics[width=12cm]{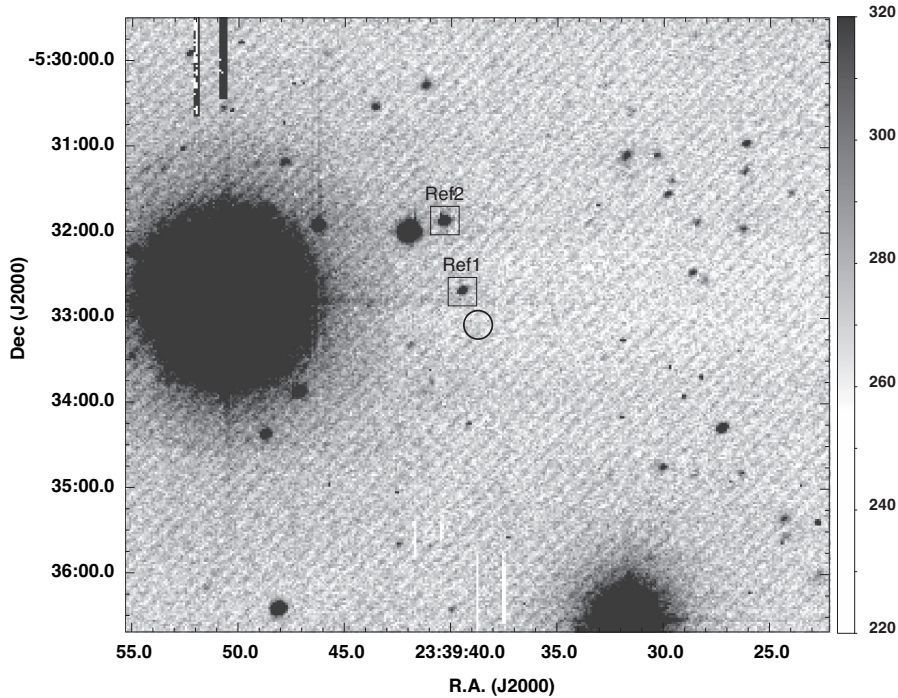} 
\caption{H$\alpha$ image of the vicinity of 2FGL J2339.6$-$0532 observed
 with the Kiso-105cm Schmidt telescope.  The circle region with a radius
 of 10\arcsec indicates the location of the target.  Two rectangular
 regions indicate the photometric reference stars shown in Table
 \ref{field_photo}.}\label{H_alpha}
  \end{center}
\end{figure*}

2FGL J2339.6$-$0532 was observed from September 22 to October 7 2011
utilizing the global telescope network, ``Optical and Infrared
Synergetic Telescopes for Education and Research (OISTER)''
\footnote{\url{http://oister.oao.nao.ac.jp/}}.  OISTER consists of 14
independent observatories that are funded by Japanese universities and
research associations.  For this work, we also asked for photometry
observation from the Kottamia Astronomical observatory which has a 188
cm reflector.  Thanks to the locations of the observatories distributed
across the globe, we can continuously trace the variability all day
long.  Moreover the telescopes cover a wide range of wavelengths from
K{\rm s} to B band which can provides phase resolved spectral energy
distribution (SED) of the target.  The conducted photometric
observations are summarized in Table \ref{OISTER_obs_summary}.  The
locations of the reference stars and the target are shown in Figure
\ref{H_alpha}.

\subsection{H$\alpha$ imaging}
In the case of PSR B1957$+$20, the binary system is surrounded by an
H$\alpha$ bow-shock due to its super sonic proper motion with respect to
the surrounding ISM \citep{1988Natur.335..801K}.  To search for a
similar feature around 2FGL J2339.6-0532, we took H$\alpha$ images with
the Kiso Schmidt telescope at the minimum phase to prevent contamination
from the companion star.  Figure \ref{H_alpha} shows the obtained
H$\alpha$ image in the vicinity of 2FGL J2339.6-0532 utilizing an
H$\alpha$ narrow band filter ($\Delta \lambda = 9$ nm), in which we
could not find the evidence of bow-shock nebula.

To evaluate the detection limit, we calculated the standard deviation of
the 33\arcsec $\times 67\arcsec$ (50$\times$100 pixel) rectangular
region near the target and obtained $STDEV = 7.82$ ADU.  The sky
background of the CCD image shows a weak striped pattern running in the
southeast-to-northwest direction, with a peak-to-peak value of $\sim 8$
ADU, which seems to be the dominant noise component restricting the
detection limit.  Regardless, if we adopt a conservative threshold with
a 3$\sigma$ confidence level, the detection limit becomes 23.5 ADU.
Based on the flux calibrations of the reference stars listed in Table
\ref{field_photo}, the conversion coefficient of ADU to energy flux,
$(8.3 \pm 0.1) \times 10^{-18}$ erg s$^{-1}$ cm$^{-2}$ ADU$^{-1}$, was
observed for the H$\alpha$ filter with a bandwidth of 9 nm.  Finally, we
obtained a 3$\sigma$ detection limit of $< 1.9\times10^{-16}$ erg
s$^{-1}$ cm$^{-2}$ pixel$^{-1}$, which corresponds to a surface
brightness of $< 8.7\times10^{-17}$ erg s$^{-1}$ cm$^{-2}$ arcsec$^{-2}$
\quad\footnote{Kiso 2k CCD's pixel resolution is 1.5\arcsec
\citep{2001PNAOJ...6...41I}.}.
\begin{deluxetable*}{lcc|c}
\tabletypesize{\scriptsize}
\tablecolumns{3}
\tablecaption{Summary of Field photometry\label{field_photo}}
\tablewidth{0pt}
\tablehead{
 \colhead{} & 
 \colhead{Reference-1} & 
 \colhead{Reference-2} &
 \colhead{A$_{\lambda}\tablenotemark{$\dagger$}$}
}
\startdata
 Name(USNO-2.0A) &
 U0825\_19993817&
 U0825\_19993871&\\
 Coordinate(J2000.0) &
 (23:39:39.487, $-$05:32:40.56)  &
 (23:39:40.366, $-$05:31:51.89)  &\\
\hline
 B &
 $ 18.435\pm0.014$&
 $ 17.603\pm0.008$&
 0.120\\
 V &
 $ 17.834\pm0.018$&
 $ 16.828\pm0.008$&
 0.091\\
 R &
 $ 17.484\pm0.016$&
 $ 16.388\pm0.007$&
 0.072\\
 I &
 $ 17.296\pm0.020$&
 $ 16.105\pm0.008$&
 0.050\\
 g' &
 $ 18.098\pm0.011$\tablenotemark{a}&
 $ 17.177\pm0.006$\tablenotemark{a}&
 0.110\\
 J &
 $ 16.47\pm0.10$\tablenotemark{b}&
 $ 15.19\pm0.04$\tablenotemark{b}&
 0.024\\
 H &
 $ 16.45\pm0.23$\tablenotemark{b}&
 $ 14.87\pm0.05$\tablenotemark{b}&
 0.015\\
 K{\rm s} &
 $ 16.12$\tablenotemark{b,c}&
 $ 14.55\pm0.08$\tablenotemark{b}&
 0.011\\
 H$\alpha$ (656 nm) &
 $314\pm46 \quad\mu$Jy\tablenotemark{d}&
 $854\pm170 \quad \mu$Jy\tablenotemark{d}&
 \nodata
\enddata
\tablecomments{Errors are with 1$\sigma$ confidence level.}
\tablenotetext{$\dagger$}{Galactic extinction at the target coordinate
\citep{2011ApJ...737..103S,1989ApJ...345..245C}}.
\tablenotetext{a}{SDSS magnitudes were calculated from B-band and V-band magnitudes based on \cite{2002AJ....123.2121S}.}
\tablenotetext{b}{For the flux calibrations of J, H, and K{\rm s}, we referred the 2MASS catalog \citep{2006AJ....131.1163S}.}
\tablenotetext{c}{The catalog error of the Reference-1 for K{\rm s}-band
was not available.}
\tablenotetext{d}{The flux density was evaluated from the absorption
 corrected SED assuming a simple balckbody model.  The estimated
 temperature of the reference stars were $T_{\rm BB,ref1} = 7320\pm160$K,
 $T_{\rm BB,ref2} = 6300\pm180$ K.}
\end{deluxetable*}

The prototype black widow PSR B1957+20, residing at a distance of 1.6
kpc from the earth, has a bow-shock nebula extending over $\sim$
30\arcsec $\times$ 30\arcsec with an H$\alpha$ flux of
$3.3\times10^{-14}$ erg s$^{-1}$ integrated over the entire nebula
\citep{1988Natur.335..801K}.  If our target is accompanied by an
equivalent bow-shock nebula, the H$\alpha$ flux and the size should be
$6.1\times10^{-14}$ erg s$^{-1}$ cm$^{-2}$ and $\sim 41\arcsec \times 41
\arcsec$, respectively.  Therefore, the expected surface brightness is
$3.6\times10^{-17}$ erg s$^{-1}$ cm$^{-2}$ arcsec$^{-2}$, which is
slightly lower than the detection limit.

Compared with the other bow-shock PWNe, the size and intensity both seem
to be scattered across a range of two orders of magnitude, possibly
reflecting their shock conditions \citep{2002ApJ...575..407C}.
Therefore, we conclude that this observation provides only a weak upper
limit of surface brightness in the H$\alpha$ band.

\subsection{Photometry}
The other telescopes except for the Kiso observatory carried out
photometric observations covering Ks$\sim$B bands from September 9th
2011 to September 30th 2011.  For flux calibration we chose two
reference stars near the target object, U0825\_19993817 (R=17.5) and
U0825\_19993871 (R=16.4) in USNO-2.0A catalog, so that all of the
observatories can employ the common references without saturation.  The
photon fluxes of the references stars were measured and compared to the
standard stars in the Landolt catalog \citep{1992AJ....104..340L} with
MSI of the Pirka telescope in Hokkaido prefecture (Japan).  Since the
Pirka employs the Johnson-Cousins filter system, we estimated the photon
flux in the SDSS g'-band based on the observed B and V magnitudes using
a conversion equation proposed by \cite{2002AJ....123.2121S},
\begin{equation}
 g' = V + 0.54 (B-V) - 0.07.
\end{equation}
For J, H, and Ks, we employed the photometric data in the Two Million
All Sky Survey (2MASS) catalog for flux calibration
\citep{2006AJ....131.1163S}.  The obtained magnitude of the reference
stars are summarized in Table \ref{field_photo}.

\begin{figure}[b]
\begin{center}
 \includegraphics[width=8.5cm]{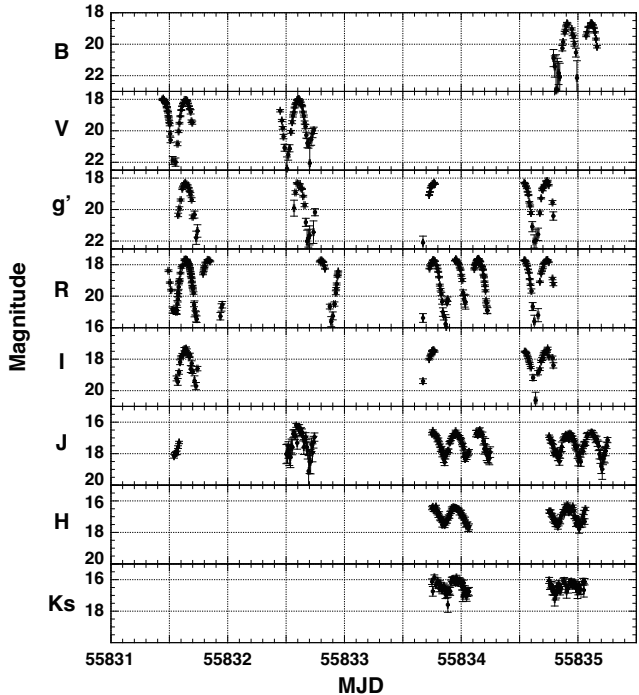}
 \caption{Multi-color light curve of 2FGL J2339.6$-$0532 obtained via OISTER.}
 \label{unfolded_lc}
\end{center}
\end{figure}

Figure \ref{unfolded_lc} shows the light curves of 2FGL J2339.6$-$0532
obtained by OISTER.  We re-confirmed the clear sinusoidal modulation as
reported by \cite{2012ApJ...747L...3K} and \cite{2011ApJ...743L..26R}.
The absolute flux seems consistent with the past observations: the
modulation amplitude is about 4.5 magnitude in the R-band and is larger
at shorter wavelength.  In R-band, we successfully observed the maximum
phases of the target 8 times during the observation campaign.  The
photon flux at the maximum phase is $R_{\rm max} = 17.7$ mag on average
and varies among peaks within a range of $\pm 0.1$ mag.  There was no
irregular activity like the flares observed in 2FGL J1311.6$-$3429
\citep{2012ApJ...757..176K}.

\subsection{Phase-resolved SED}
To discuss the energetics quantitatively, we estimated the energy flux
based on the photometric data shown in Figure \ref{unfolded_lc}.  First
we corrected the galactic extinction at the coordinate of the target
\citep{2011ApJ...737..103S, 1989ApJ...345..245C}\footnote{In the above
extinction evaluation we utilized
\url{http://ned.ipac.caltech.edu/forms/calculator.html}, and
\url{http://dogwood.physics.mcmaster.ca/Acurve.html}}; the employed
extinction at various wavelength are listed in Table \ref{field_photo}.
Then we converted the absorption corrected magnitudes to energy flux
using conversion equations presented in \citet{1996AJ....111.1748F} and
\citet{2005PASP..117..421T} for optical (B, V, R, I, g') and IR(J, H,
K{\rm s}) energy bands, respectively.

\begin{figure*}[t]
 \begin{center}
\includegraphics[width=17cm]{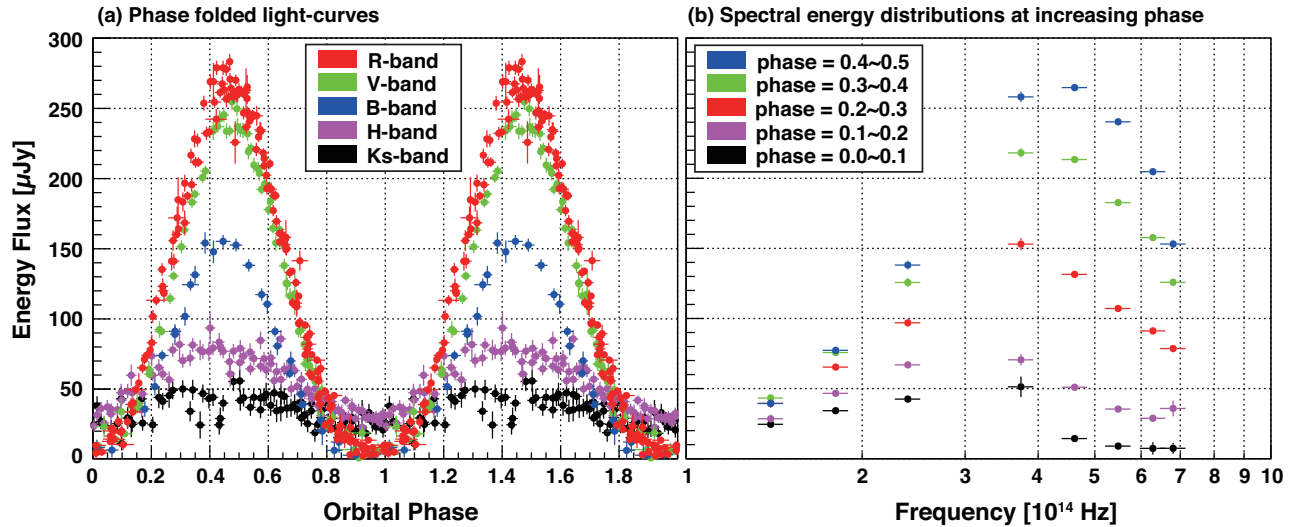} \caption{(a)Phase
folded multi-color light curves.  The assumed orbital period is 4.63435
hr and Phase=0 is set to be MJD=55500 day.  Marker colors represent the
energy bands: red= R, green= V, blue=B, magenta=H, and black=K{\rm s}.
(b)Spectral energy distributions at increasing phase from the minimum to
the maximum.  Colors correspond to the orbital phase: black=0-0.1,
magenta=0.1-0.2, red=0.2-0.3, green=0.3-0.4, and blue=0.4-0.5.}
\label{SED}
 \end{center}
\end{figure*}

Figure \ref{SED}---(a) shows the yielded energy fluxes as functions of
the binary orbital phase.  Phase=0 was set to be MJD=55500.0.  While the
optical light-curves show symmetric structures, IR (H and K{\rm s})
light-curves possess somewhat asymmetric shapes.  This may be due to the
geometry of the companion star or evaporating stellar gas.  Panel (b)
shows the spectral energy distribution during the brightening phase from
orbital phase= 0 to 0.5 in panel (a).  Clearly the peak frequency
increases as the orbital phase increases from orbital phase 0 to 0.5,
implying that the temperature of the companion is increasing with the
orbital phase.  At the maximum phase the SED is fitted with a black-body
model with an effective temperature of $T_{\rm eff} = 7540\pm130$ K,
which is consistent with the past study \citep{2011ApJ...743L..26R}.

\begin{figure}[t]
 \includegraphics[width=8.5cm]{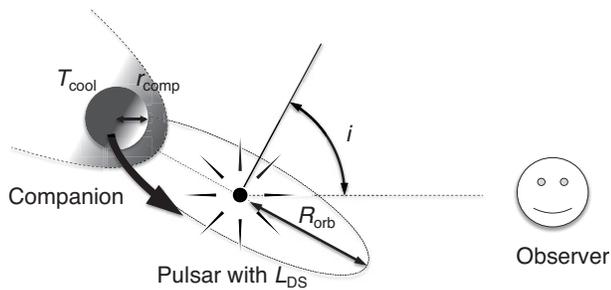} 
 \caption{Schematic image of the assumed emission model in which a
 companion star with a radius $r_{\rm comp}$ is orbiting around a pulsar
 at an orbital radius $R_{\rm orb}$.  The hemisphere of the companion
 star facing the pulsar is heated.  In this paper, the inclination
 angle $i$ is defined as the angle between the line of sight and the
 normal line to the orbital plane.}  \label{schematic_image}
\end{figure}

In this paper, we assumed a simple emission model from the surface of
the companion star to explain the observed optical emission.  As
described in Figure \ref{schematic_image}, we supposed that a companion
star with a radius $r_{\rm comp}$ is orbiting around a pulsar at an
orbital radius $R_{\rm orb}$, and the hemisphere of the companion that
faces the pulsar is heated by the pulsar wind, while the opposite side
of the companion has ordinal temperature $T_{\rm cool}$.  Note that this
model does not take into account the effect of energy transfer via
convection or advection on the surface of the companion, therefore the
temperature distribution of the companion is simply described by the
energy injection via pulsar wind although the heating efficiency is
unclear.  In this paper, we assumed that the spin-down energy is
perfectly converted into the isotropic pulsar wind and the injected
energy via the pulsar wind heats the companion up with a heating
efficiency of $f$ for simplicity.  We also adopted a spin-down
luminosity of $2.3\times10^{34}$ erg s$^{-1}$ based on the timing
analysis in radio band recently reported by \cite{2014AAS...22314007R}.

\begin{figure}[h]
 \includegraphics[width=8.5cm]{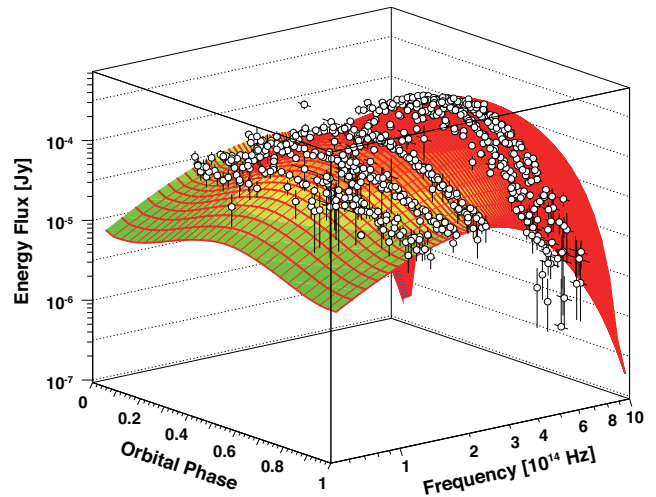}
 \caption{Phase-resolved SED fitted with a model function.}
 \label{3D_SED_fit}
\end{figure}

\begin{figure}[t]
 \includegraphics[width=8.5cm]{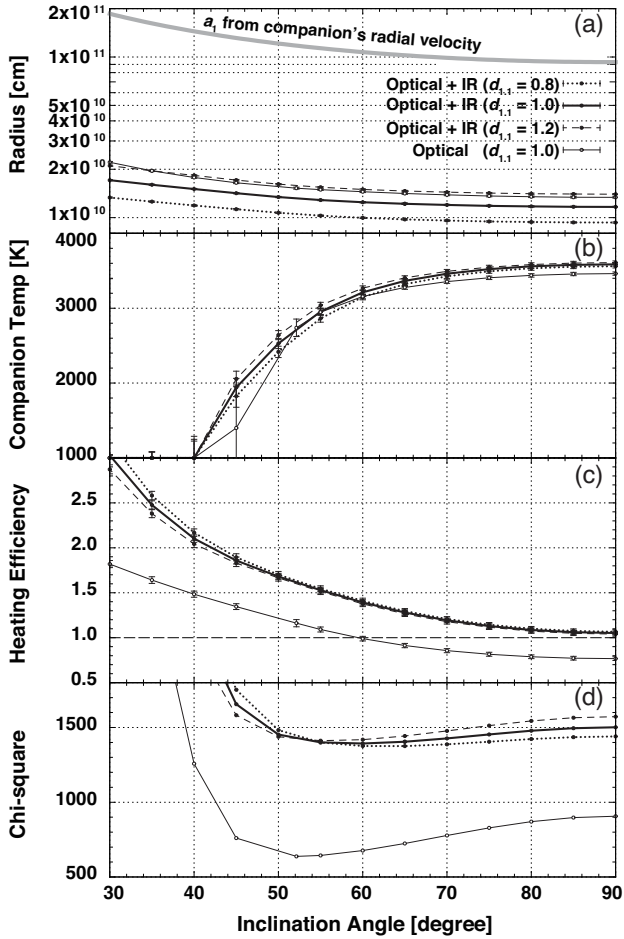} \caption{The
 obtained parameters from model-fitting of the phase-resolved SED as
 functions of inclination angle$i$.  Panel (a)$\sim$(d) show the orbital
 radius $R_{\rm orb}$ and companion radius $r_{\rm comp}$, the companion
 temperature at the cool side, the heating efficiency $f$, and the
 resultant $\chi^{2}$ (d.o.f = 901), respectively.  The gray-bold line on
 panel (a) represents the semi-major axis of the companion calculated
 from the radial velocity of 350 km s$^{-1}$ with respect to the mass
 centroid of the system \citep{2011ApJ...743L..26R}.  In this
 calculation we employed a spin-down luminosity of the pulsar, $L_{\rm
 SD} = 2.3 \times 10^{34}$, based on \cite{2014AAS...22314007R}.}
 \label{param_vs_inclination}
\end{figure}

\begin{figure}[t]
 \includegraphics[width=8.5cm]{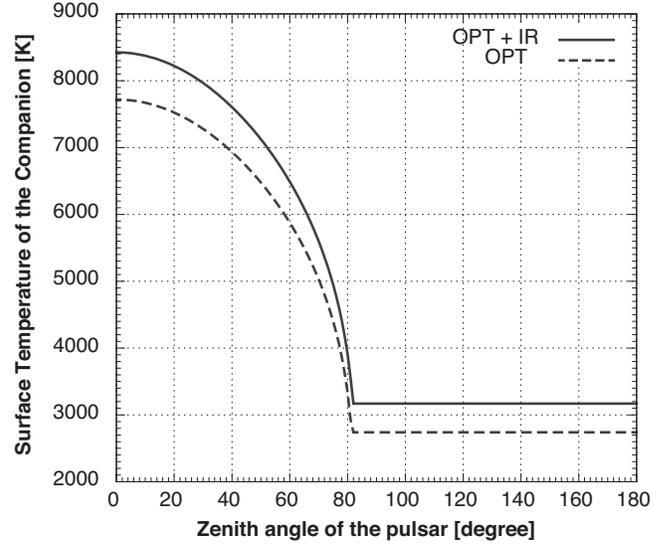}
 \caption{Simulated temperature of the companion surface as a function
 of zenith angle of the pulsar from the position for which the
 calculation is performed.  The simulations were performed for
 the best-fit parameters listed in Table \ref{SEDfit_results}.}
\label{multicolor_temp}
\end{figure}

We then fitted the obtained phase-resolved SED with the model function
described above (Figure \ref{3D_SED_fit}).  Since the model seems to
correlate weakly with the inclination angle, we fixed the inclination
angle $i$.  The resultant model parameters are summarized in Figure
\ref{param_vs_inclination} as functions of inclination angle.  Although
the fit seems rather poor --- mainly because of the large residual in
the IR band --- we found that an inclination angle $i = 59\degr$
minimizes the $\chi^2$.  For comparison, we also attempted this analysis
without IR data and obtained a best-fit parameter of $i = 52\degr$.
These results seem consistent with the inclination angle reported by
\citet{2011ApJ...743L..26R}.  The obtained best-fit parameters are
summarized in Table \ref{SEDfit_results}.  Additionally, the obtained
temperature distributions of the companion surface are plotted in Figure
\ref{multicolor_temp} as functions of the zenith angle of the pulsar
from the measured point on the companion surface.

\begin{deluxetable*}{lcc}
\tabletypesize{\scriptsize}
\tablecolumns{3}
\tablecaption{Obtained parameters of the binary system from the phase-resolved SED\label{SEDfit_results}}
\tablewidth{0pt}
\tablehead{
 \colhead{Data set}& 
 \colhead{Optical+IR\tablenotemark{a}}&
 \colhead{Optical\tablenotemark{a}}
}
\startdata
 Orbital radius ($R_{\rm orb}$)\tablenotemark{b}&
 $1.08 \pm 0.02 \quad \times 10^{11}$ cm&
 $1.18 \pm 0.02 \quad \times 10^{11}$ cm\\
 Companion radius ($r_{\rm comp}$) &
 $1.26 \pm 0.02 \quad \times 10^{10} \; d_{\rm 1.1}$ cm\tablenotemark{c}&
 $1.53 \pm 0.02 \quad \times 10^{10} \; d_{\rm 1.1}$ cm\tablenotemark{c}\\
 Companion Temperature ($T_{\rm cool}$)&
 $3170 \pm  70  \quad$ K&
 $2740 \pm 120  \quad$ K\\
 Heating Efficiency ($f$)\tablenotemark{d}&
 $1.42 \pm 0.05$&
 $1.16 \pm 0.04$\\
 Inclination angle ($i$)&
 $59.\degr0 \pm 1.\degr5$&
 $52.\degr1 \pm 1.\degr0$\\
 Bolometric Luminosity\tablenotemark{e}&
 $1.09 \pm 0.17\quad \times10^{32}\; d_{\rm 1.1}^2$ erg s$^{-1}$&
 $1.12 \pm 0.16\quad \times10^{32}\; d_{\rm 1.1}^2$ erg s$^{-1}$\\
 $\chi^2$ & 1392.6& 637.6\\
\enddata
\tablecomments{Errors are with 1 $\sigma$ confidence level.}  
\tablenotetext{a}{``Optical'' and ``IR'' correspond to [B, V, R, I, g']
 data-set and  [J, H, Ks] data-set, respectively.}  
\tablenotetext{b}{$R_{\rm orb}$ are calculated based on radial velocity
 of the companion star reported by \cite{2011ApJ...743L..26R}.}  
\tablenotetext{c}{$d_{\rm 1.1}$ is the distance in unit of 1.1 kpc.}  
\tablenotetext{d}{Assuming an isotropic pulsar wind with a luminosity of
 $2.3\times10^{34}$ erg s$^{-1}$ based on \cite{2014AAS...22314007R}.}
\tablenotetext{e}{Bolometric luminosity of the companion's heated hemisphere.}
\end{deluxetable*}

\section{X-ray observation}
\subsection{Observation}
2FGL J2339.6$-$05312 was observed with the {\itshape Suzaku} satellite
on Jun. 29th, 2011.  The target was observed at the XIS nominal point
with the normal imaging mode.  The net exposure time after the standard
data reduction process was 96 ks.  The accumulated photons are 2555
counts with the front-illuminated CCDs (XIS0 + XIS3) and 1874 counts
with the back-illuminated CCD (XIS1) in the 0.4$-$8 keV energy band.
During the observation, the event rates were $\sim 8.25 \pm 0.28 \times
10^{-3}$ (XIS0, XIS3) photons s$^{-1}$ and $8.89 \pm 0.50 \times
10^{-3}$ photons s$^{-1}$ (XIS1) in the energy band.

\subsection{Spectral analysis}
\begin{figure}[t]
 \includegraphics[width=8.5cm]{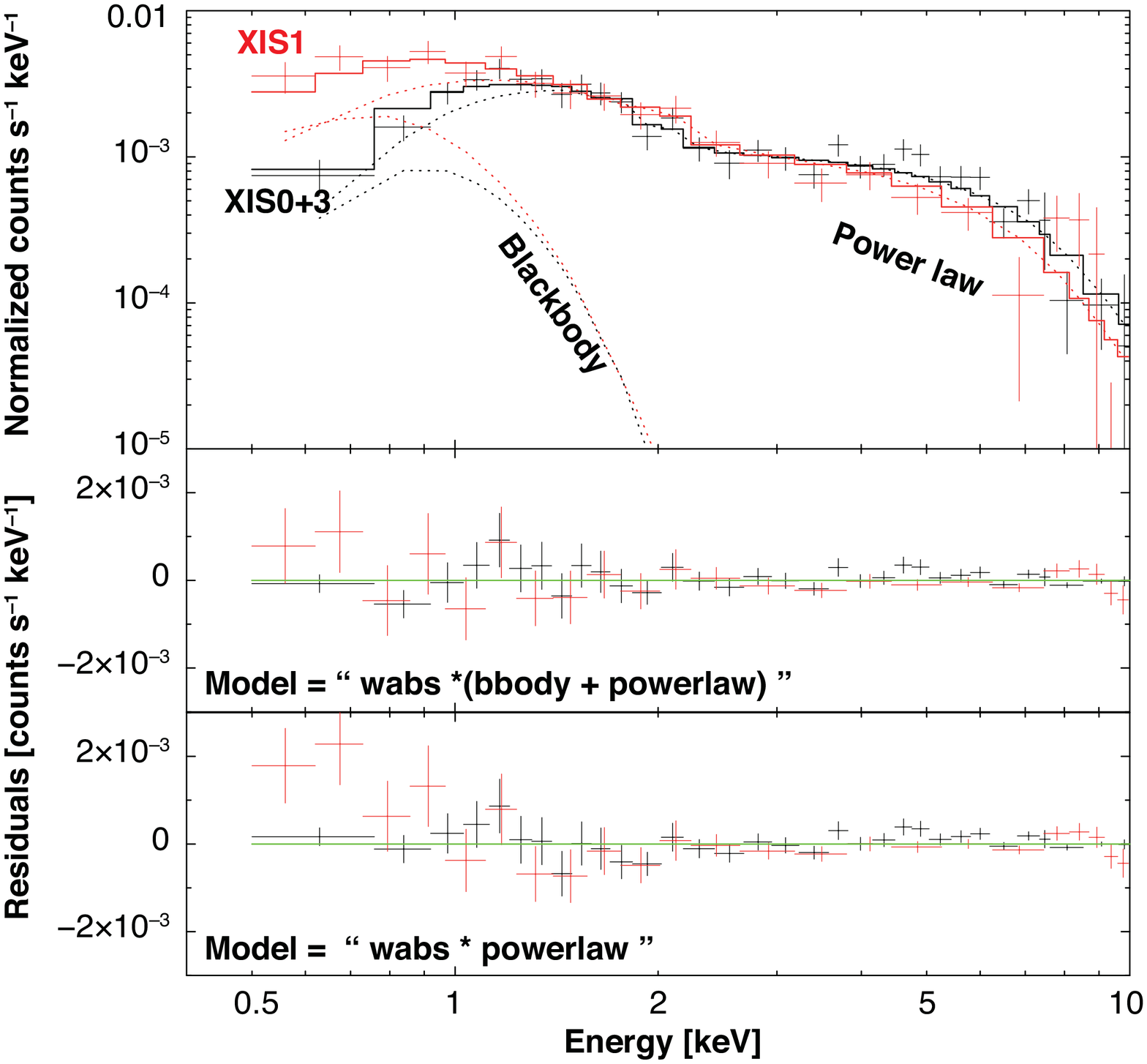} \caption{X-ray
 spectra observed by the XIS aboard the Suzaku satellite.  Black and
 red data points corresponds to the spectra observed with the front
 illuminated CCDs (XIS0 and XIS3), and with the back illuminated CCD
 (XIS1), respectively.  Dotted-lines show the best-fit model functions
 consisting of a low-temperature black-body and a power-law function.  The
 model parameters are found in Table \ref{spec_results}.} 
\label{Xspec}
\end{figure}

Figure \ref{Xspec} shows the X-ray spectrum obtained with the Suzaku.
At first, we tried to fit the X-ray spectra with a single power-law
function however the resultant fit was poor ($\chi^2/\nu$ = 67.71/55)
due to residuals below 1 keV, implying the presence of a soft excess
component.  We therefore added a black-body to the model function for
the soft excess, and obtained a better fit ($\chi^2/\nu = 51.82/55$).
Dotted lines on Figure \ref{Xspec} describe the best-fit model
functions.  The resultant parameters of the model fitting are summarized
in Table \ref{spec_results}.  In both case, the column density converged
on zero.  The resultant upper limit is $<1.2\times10^{21}$ cm$^{-2}$
with 90 \% confidence level, and is consistent with the Galactic
absorption of $3.23\times10^{20}$ cm$^2$ for this direction
\citep{1990ARA&A..28..215D,2005A&A...440..775K}.
\begin{deluxetable*}{lcc}
\tabletypesize{\scriptsize}
\tablecolumns{3}
\tablecaption{Results of model fitting of the averaged X-ray spectra\label{spec_results}
}
\tablewidth{0pt}
\tablehead{
 \colhead{Model} & 
 \colhead{wabs*(bbody + power-law)}&
 \colhead{wabs*power-law}
}
\startdata
$N_{\rm H}$ ($10^{21}$ cm$^{-2}$) &
$ < 1.2$ (0) &  $ < 0.13$ (0) \\
Blackbody Temperature (keV)&
 $0.15 \pm 0.06$  & \nodata \\
Blackbody Radius (km)\tablenotemark{$\dagger$} &
 $0.28^{+0.93}_{-0.16}\quad$ & \nodata\\
Power-law Photon Index&
 $1.14^{+0.14}_{-0.15}$ & $1.32\pm0.08$ \\ 
Power-law Flux$_{\rm 0.5-10keV}$ ($10^{-13}$ erg cm$^{-2}$ s$^{-1}$)&
 $2.50 \pm 0.16$ & $2.50 \pm 0.15$\\
Total flux$_{\rm 0.5-10keV}$ ($10^{-13}$erg cm$^{-2}$ s$^{-1}$)&
 $2.67^{+0.01}_{-1.25}$&  $2.46^{+0.04}_{-0.73}$\\
 $\chi^2$ (dof)&
 51.82 (51) & 67.71(53)\\
\enddata
\tablecomments{Errors are with 90 \% confidence level.}
\tablenotetext{$\dagger$}{Blackbody radius assuming a distance of 1.1 kpc.}
\end{deluxetable*}

\begin{figure*}[t]
\begin{center}
\includegraphics[height=10cm,width=15cm]{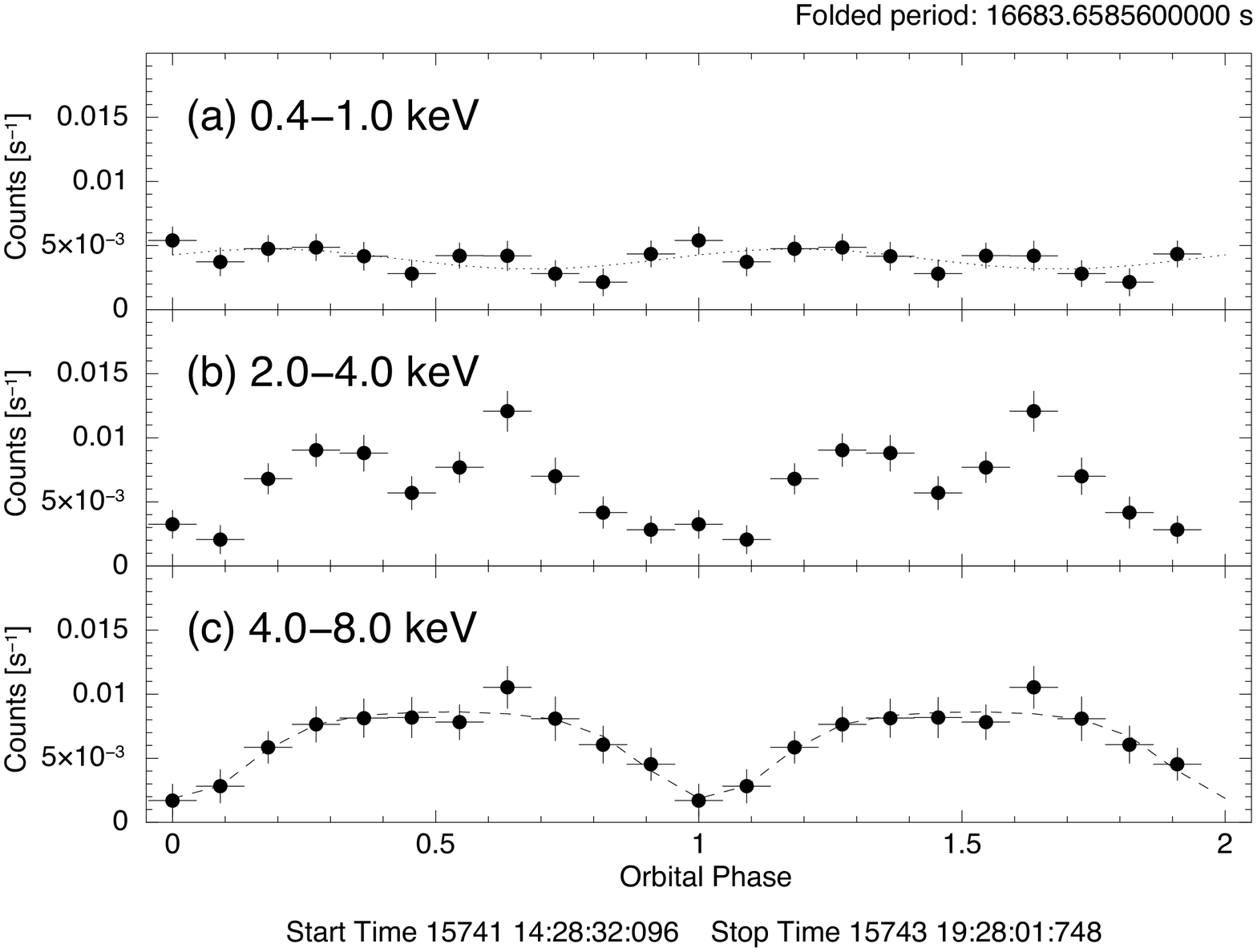}
\caption{Phase-folded X-ray light curves for (a)0.4$-$1.0 keV
(b)2.0$-$4.0 keV and (c)4.0$-$8.0 keV energy bands.  Phase=0 is set to
be MJD=55500.0 in the same way as the optical light curves.  Dotted line
in panel (a) is the best-fit sinusoidal curve for evaluating the
modulation factor.  Dashed line in panel (c) shows a model function for
synchrotron emission from a thin layer covering the companion
hemisphere.}  \label{folded-XLC}
\end{center}
\end{figure*}

In order to clarify the origin of these two components, phase-folded
X-ray light curves were generated as shown in Figure \ref{folded-XLC}.
In this figure, Phase=0 was set to be MJD=55500.0 in the same way as the
optical light curves shown in Figure \ref{SED}$-$(a).  Panel
(a)$\sim$(c) in Figure \ref{folded-XLC} represent the energy bands
0.4$-$1.0 keV, 2.0$-$4.0 keV and 4.0$-$8.0 keV, respectively.  Clearly,
the hard X-ray light curves show clear modulation coinciding with the
orbital motion.  This timing coincidence may imply that the non-thermal
emission originates from the companion surface.  However the shape of
the light curves is somewhat different from the sinusoidal shapes
observed in optical, i.e., a double-peaked shape at 2.0$-$4.0 keV band,
and a flat top shape at 4.0$-$8.0 keV.  On the other hand, the soft
X-rays seem to be steady.  This may imply that the soft component has a
different emitting region, namely the pulsar.

To evaluate the upper limit of the modulation factor in the soft X-ray
energy band, we tried to fit the X-ray light curve with a sinusoidal
curve and a constant component, $A \sin (2\pi x) + B$, where $x$ is the
orbital phase.  The obtained amplitude of the sinusoidal curve is $A =
(0.80\pm 0.34)\times 10^{-3}$ counts s$^{-1}$ against the constant
component of $B = (3.95\pm 0.23)\times 10^{-3}$ counts s$^{-1}$.
Therefore, the upper limit of the modulation factor is $A/B < 28.6$\%
with 1 $\sigma$ confidence level.  Note that the maximum point was not
fixed to be phase=0.5 in this estimation; this helped to maximize the
modulation factor.  The best-fit model curve is shown in Figure
\ref{folded-XLC} (a).

If we adopt the two-component model consisting of a blackbody and a
power-law function from the X-ray spectroscopy, we can roughly evaluate
the contributions from these two components to the constant emission.
The phase-averaged absorbed-photon flux for the blackbody and the power
law in the 0.4$-$1.0 keV energy band are both $2.0\times10^{-5}$ photons
s$^{-1}$.  Assuming that the constant component originates in the
blackbody, the modulation factor is expected to be $(PL)/(BB + PL)\sim
50$ \%, which is about twice of the measured value.  Taking into account
the large errors arising from the uncertainties of the blackbody
components (see Table \ref{spec_results}), the measured modulation
factor should be larger because of the tight upper limits to the
blackbody component.  This may indicate that the constant emission may
consist of the blackbody from the NS surface and the non-thermal
component originating in the pulsar magnetosphere.

\subsection{Temporal analysis}
\begin{figure*}[t]
 \begin{center}
  \includegraphics[width=15cm]{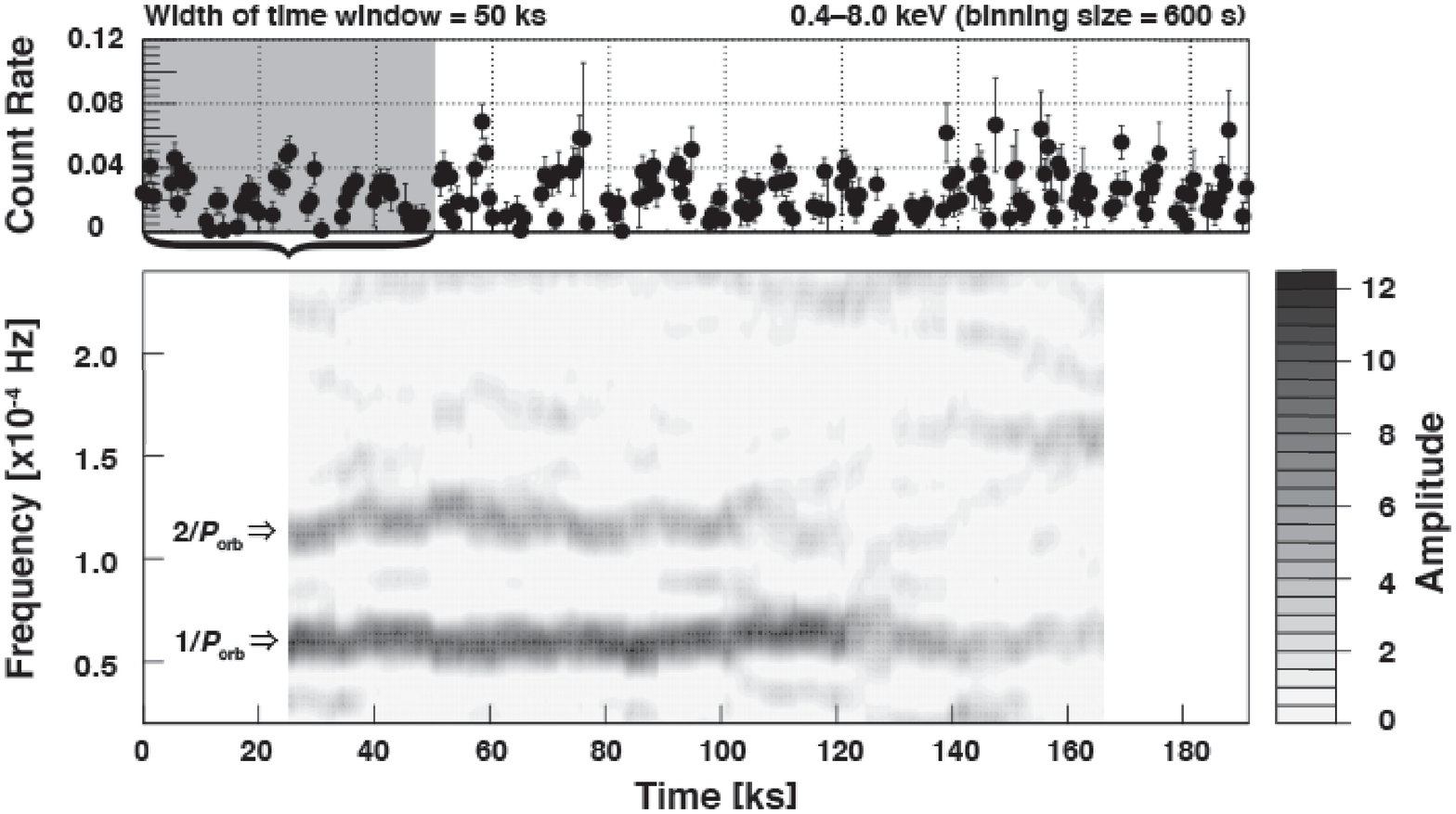} 
  \caption{Upper panel shows the X-ray light curve for energy band of
  0.4$-$8.0 keV with a bin size of 600 s.  Lower panel shows the running
  periodogram generated from the X-ray light curve shown in the upper
  panel with the Lomb-Scargle algorithm.  We employed a 50ks rectangular
  window function that scans over the entire X-ray light curve of 190 ks.
  The horizontal line in the lower panel corresponds to the center time of
  the sampled time span of 50 ks.  The gray scale shows arbitrary
  amplitude of modulation.  Note that the amplitude of 6.3 correspond to
  90 \% confidence level.}  \label{fig:Lomb}
  \end{center}
\end{figure*}

The upper panel of Figure \ref{fig:Lomb} shows the obtained X-ray light
curve with a bin size of 600 s for the energy range of 0.4$-$8.0 keV.
Note that the whole length of the light curve is 190 ks, containing
occultation periods of earth, and corresponding to the twice of the net
exposure time of 96 ks.  Although the photon statistics seem poor, the
X-rays show periodic variability coincident with an orbital period of
16683 s.  We expected to observe the flare-like features reported in
previous studies \citep{2012ApJ...757..176K, 2013Natur.501..517P,
2014A&A...567A..77F, 2014MNRAS.438..251L}, which may be related to the
accretion activity; however, no significant evidence for this was
discovered.  Nonetheless, the shapes of the light curve seem to change
over time.

In order to clarify the temporal variation in the waveform of the X-ray
light curve, we traced the temporal changes of the power spectrum with
the Lomb-Scargle algorithm, which can be applied to an unevenly sampled
data set.  In this analysis, we employed a rectangular function as
a window function to detect small changes from a short data sequence.
The width of window function was set to 50 ks in order to cover three
orbital laps and to detect the light-curve variations for the total
exposure time of 190 ks.  We then scanned the light curve by sweeping
the window function, and generated a power spectrum for every timespan
in the light curve.  Note that this method cannot detect rapid
variations of the power spectra (faster than a few tens of
kilo-seconds), because the width of the time window has a finite
timespan.

The lower panel of Figure \ref{fig:Lomb} shows the obtained periodogram
as a function of time, which represents the central time of the sampled
time window.  Therefore, the periodogram at $t = 25$ ks consists of the
X-ray light curve in the time span of $0\leq t \leq 50$ ks.  Both ends
of the grayscale map are omitted because the time window can sweep in
the range of the obtained light curve.  The sweeping step size was set
to 600 s.  Therefore, the time resolution of the running periodogram is
about 230 which is much greater than the timescale that can be detected
by this method.  Nevertheless, a clear peak structure was detected at a
frequency of around $f_1 = 0.6\times 10^{-5}$ Hz, which corresponds to
the orbital frequency.  On the other hand, $f_2 = 1.2 \times 10^{-5}$ Hz
appears to denote the second harmonic of the orbital frequency, which
may be related to the dip structure of the double-peaked light curve.
Interestingly, these two components in the power spectra decrease
sequentially after $t$ = 1.2$\times 10^{4}$ s.

\section{Discussion}
\subsection{Constraints on the inclination angle of the binary orbit}
The model fitting of the phase-resolved SED resulted in inclination
angles of $i=59\degr$ from the entire data set (covering optical + IR
bands), and $i=52\degr$ from the optical-only data.  This discrepancy
seems to arise from the systematic errors of the reference stars in the
IR band (J, H, Ks), resulting in an underestimate of the target
luminosity in the IR band.  Therefore, the SED of the optical+IR data
set tended to return higher companion temperatures than the optical-only
data.  On the other hand, it is also difficult to constrain the
temperature with the optical-only data (B, V, g', R, and I), which may
then result in an underestimate of the companion temperature.
Therefore, the true inclination angle must be in the range of $52\degr <
i <59\degr$.  Hereafter, we take $i\sim 55\degr \pm 5\degr$.

We should also summarize the pulsar parameters in preparation for the
discussion below.  \citet{2014AAS...22314007R} recently reported the
detection of radio and gamma-ray pulse emission from the central MSP
with a period of 2.88 ms with a spin-down luminosity of $L_{\rm SD} =
2.3\times10^{34}$ erg s$^{-1}$.  Based on these parameters, the pulsar's
surface magnetic field can be estimated, assuming that the pulsar's
rotation energy is dissipated by dipole radiation
\citep{1994hea..book.....L},
\begin{equation}
 B_{\rm *} = 3.0 \times 10^{19}(P\dot{P})^{1/2} = 
  1.9 \times 10^8 I_{45}^{-1/2} \quad \mbox{G},
\end{equation}
where $I_{\rm 45}$ is the pulsar's moment of inertia in units of
$10^{45}$ g cm$^2$, which is a typical magnetic field for a standard
MSP.

\subsection{Condition for mass Accretion}
In the case of neutron star (NS) binary systems, the mass accretion
process is thought to be controlled by the propeller effect
\citep{1986ApJ...308..669S}, e.i., the pressure balance between the ram
pressure of accreting material $p_{\rm ram}$ and the magnetic pressure
of the pulsar $p_{\rm B}$.  Here we define $r_{\rm m}$ where $p_{\rm B}$
and $p_{\rm ram}$ are in balance,
\begin{eqnarray}
 &\frac{B_{\rm *}^2}{8 \pi}
  \left(\frac{r_{\rm *}}{r_{\rm m}} \right)^6 =
  \frac{\dot m}{4\pi r_{\rm m}^2} 
  \left(\frac{2GM_{\rm *}}{r_{\rm m}}\right)^{1/2},\nonumber&\\
 &r_{\rm m} = 
  \left(\frac{B_{\rm *}^4 r_{\rm *}^{12}}
   {8GM_{\rm *}\dot{m}^2}\right)^{1/7}\quad\quad\quad\quad&\nonumber\\
 &= 2.5\times 10^6 
  \left(\frac{B_{\rm *}}{10^8\mbox{ G}} \right)^{4/7}
  \left(\frac{r_{\rm *}}{10 \mbox{ km}} \right)^{12/7}&\nonumber\\
 &\left(\frac{M_{\rm *}}{1.4 M_{\sun}} \right)^{-1/7}
  \left(\frac{\dot{m}}{10^{16}\mbox{ g s}^{-1}} \right)^{-2/7}
  \mbox{cm},&
\end{eqnarray}
where $G$ is the gravitational constant, $r_{\rm *}$ is the NS radius,
$M_{\rm *}$ is the NS mass, $\dot{m}$ is the mass accretion rate.
Nearby the pulsar, charged particles are trapped by the pulsar
magnetosphere and are co-rotating with the pulsar.  If the accreting gas
approaches the neutron star within a radius,
\begin{eqnarray}
 &r_{\rm c} = 
  \left( \frac{G M_{\rm*}}{\omega_{\rm *}^2} \right)^{1/3}\quad\quad\quad&\nonumber\\
 & = 1.7 \times 10^6 
  \left( \frac{M_{\rm *}}{1.4 M_{\sun}  } \right)^{1/3}
  \left( \frac{P        }{2.88\mbox{ ms}} \right)^{2/3} \quad \mbox{cm}&
\end{eqnarray}
at which the centrifugal force and the gravitational attraction are
equal, the gas will accrete on the neutron star (where $\omega_{\rm *}$
is the angular velocity of the NS).

Utilizing $B_{\rm *}$ from the pulse profile, the lower limit of the
accretion rate $\dot{m}$ that fulfills the condition of $r_{\rm c} >
r_{\rm m}$ can be calculated\footnote{In which we assumed a relation
$I_{45}\propto M_{\rm *} r_{\rm *}^2$.},
\begin{eqnarray}
 &\dot{m} > \frac{B_{\rm *}^2 r_{\rm *}^6 
  \omega_{\rm *}^{7/3}}{2^{3/2}G^{5/3}M_{\rm *}^{\rm 5/3}}\quad\quad\quad\quad&\nonumber\\
 &= 1.3\times 10^{16} 
  \left(\frac{r_{\rm *}}{10\mbox{ km}  }\right)^{8} 
  \left(\frac{M_{\rm *}}{1.4 M_{\sun}  }\right)^{-2/3} 
  \mbox{g s}^{-1}.&  \label{mdot_rate}
\end{eqnarray}

Recently, \cite{2014AAS...22314007R} reported temporal extinctions of
the pulse emission in radio implying that the accretion on the MSP is
ongoing.  If that is the case, the mass accretion rate must be higher
than Eq. \ref{mdot_rate} and the expected luminosity during the mass
accretion will amount to $2.4\times 10^{36}$ erg s$^{-1}$ which is about
5 orders of magnitude higher than the X-ray luminosity at the 0.5$-$10.0
keV energy band.  \cite{2013Natur.501..517P} reported intriguing
activities of IGR J18245$-$2452 in the intermediate stage between the
rotation and accretion power.  \cite{2014MNRAS.438..251L} and
\cite{2014A&A...567A..77F} reported weak accretion flows (below
$\dot{m}<10^{6}$ g s$^{-1}$) in IGR J18245$-$2452; these exhibit a
variety of activities possibly reflecting accretion conditions.  In case
of $r_{\rm m} \approx r_{\rm c}$, namely the weak-propeller regime,
accretion flow is only partly inhibited and the X-ray luminosity is in
the range of $10^{35} \sim 10^{37}$ erg s$^{-1}$.  For cases of $r_{\rm
m} > r_{\rm c}$, namely the strong propeller regime, the X-ray
luminosity is in the range of $10^{33} \sim 10^{34}$ with a very fast
and striking variability.  Regardless, the X-ray luminosity of 2FGL
J2339.6$-$0532 is still lower than that of IGR J18245$-$2452 in the
strong-propeller regime by an order of magnitude.

In addition IGR J18245$-$2452 exhibited another quiescent state in 2002
with an X-ray luminosity of $\sim 10^{32}$ erg s$^{-1}$, which is
comparable to the observed X-ray luminosity of 2FGL J2339.6$-$0532.
\cite{2014A&A...567A..77F} claimed that $r_{\rm m}$ reached the
light-cylinder radius in this regime.  This condition would allow for
the existence of pulsar wind and would also prevent mass accretion.
Thus, we conclude that the binary system can be in the same condition
seen in IGR J18245$-$2452 in 2002 or in the late stage of the evolution
process without mass accretion.  In the latter case, where the mass
accretion already ended, the radio-pulse extinction can also be
explained by the unstable activity of the pulsar's magnetosphere itself,
as with ``nulling'' pulsars \citep[see also references
therein.]{2012MNRAS.424.1197G}.

\subsection{Origin of the X-ray emission}
As shown in the spectral analysis, the soft X-rays below 1 keV seem to
originate from a black-body.  The obtained radius of 0.28 km for $d =
1.1$ kpc is unusually small for an accretion disc and thus the radiation
can be interpreted as the X-ray radiation from the neutron star surface.
A radius of 0.28 km is still small for a neutron star, therefore the
emitting region could be a hot spot around the polar cap area.  In order
to constrain the geometry of the polar cap, a high time resolution X-ray
observation is required.  In addition, the small variability and the
lack of eclipse in X-ray light curve are consistent with the estimated
inclination angle of $i\sim55\degr$.

In contrast, the hard X-ray light curves vary with the orbital motion.
The X-ray luminosity of the non-thermal component from the averaged
X-ray spectra amounts to $L_{\rm pow} = 3.6\times 10^{31}$ erg s$^{-1}$
for a distance of 1.1 kpc at 0.5$-$10.0 keV.  Therefore the ratio of
$L_{\rm pow}$ against the spin-down luminosity is about 0.15 \%, which
is the typical value for a standard radio pulsar
\citep{1997A&A...326..682B, 2008AIPC..983..171K}.

To explain the flat top modulation of the 4$-$8 keV light curve, two
possible emitting regions can be proposed.  One is an extended emitting
region just around the pulsar like a halo.  In this hypothesis the
minimum phase at the inferior conjunction can be explained by an eclipse
of the halo.  The observed smooth modulation requires the emitting
region to be larger than the companion, $10^{10}$cm$\sim$, which is
still small for a standard pulsar wind nebula.  In many case the
non-thermal X-rays from a pulsar wind nebula is from the downstream of
the termination shock and the radius of the termination shock is
constrained by the pressure balance.  In this case, the upper limit of
the emitter size is the orbital radius of $R_{\rm orb} \sim 10^{11}$ cm.
At this distance however the wind pressure is still too strong to be
terminated.

\begin{figure}[h]
 \begin{center}
  \includegraphics[width=8.5cm]{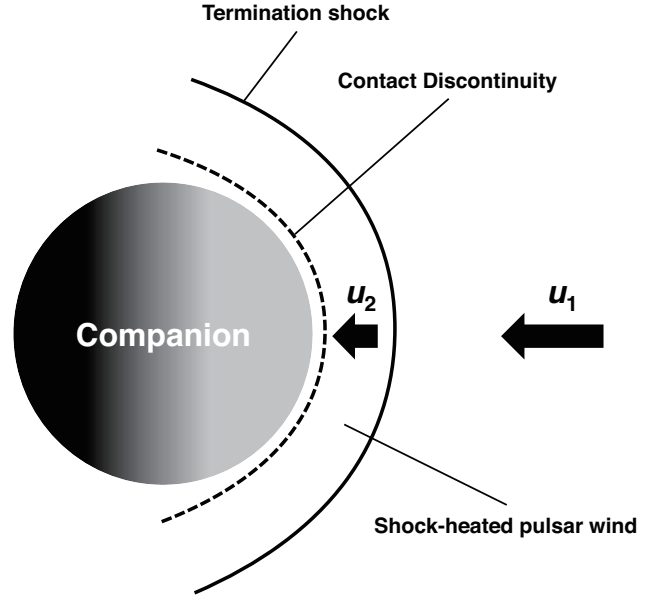}
  \caption{Schematic view of the Hard X-ray emission from the surface of
  the companion.}
  \label{schem_H-X}
 \end{center}
\end{figure}

The other hypothesis for the origin of the hard X-rays is that of a
shocked pulsar wind interacting with the companion star.  The companion
star orbiting nearby the pulsar can partially terminate the strong
pulsar wind.  In this case the flat-top light curve can be explained by
the Lorentz boost of shocked plasma.  As shown in Figure
\ref{schem_H-X}, we assumed a thin emitting layer covering the
hemisphere of the companion star.  The surface of the emitting region is
the wind termination shock and the injected electron plasma decelerated
at the shock to a bulk velocity $\beta_2$ reflecting the magnetization
parameter upstream of the shock.  At the superior conjunction phase the
shocked plasma is flowing in a direction away from us, therefore the
synchrotron luminosity may be weakened \citep{1987ApJ...319..416P}.  The
dashed line in the bottom panel of Figure \ref{folded-XLC} shows the
calculated photon flux for the thin layer of the shocked pulsar wind on
the surface of the companion with a photon index of $p = 1.14$ and an
inclination angle of $i=55\degr$.  The best-fit bulk velocity of the
shocked wind is $\beta_2 = 0.43\pm0.15$ in units of the light speed.  If
that is the case, the X-ray flux is attenuated by a factor of
\begin{equation}
\left(\frac{\sqrt{1-\beta_2^2}}{1-\beta_2 \cos (\pi/2 + i)}
\right)^{1+p} \sim 0.42.
\end{equation}

Here we employed a magnetization parameter $\sigma$ for the pulsar wind
upstream of the shock, $\sigma = B_1^2/(4\pi n_1 u_1 \gamma_1, m_e c^2)$
\citep{1984ApJ...283..694K}, where $B_1$ is the magnetic field just
upstream of the shock, $n_1$ is the number density of the pulsar wind,
$u_1$ is the flow velocity defined by $1 + u_1^2 = \Gamma_1^2$
($\Gamma_1$ is the bulk Lorentz factor of the wind upstream of the
shock), and $c$ is the light velocity.  Assuming that half of the wind
energy is carried as Poynting flux ($\sigma = 1$), $B_1$ amounts to
\begin{equation}
 B_1 = \left(\frac{\sigma}{1+\sigma}
	\frac{L_{\rm SD}}{R_{\rm orb}^2 c}\right)^{1/2} 
 = 5.5 \quad \mbox{G}.
\end{equation}
Furthermore, the pulsar wind is compressed at the shock, therefore the
downstream magnetic field $B_2$ is higher than $B_1$.  Adopting $B_2 >
5.5$ G as a lower limit, we can constrain the gyration radius of the
synchrotron electrons emitting 1 keV X-rays,
\begin{equation}
 r_{\rm gyro} = \frac{\gamma_2 m_e c^2}{eB} 
  < 3.5 \times 10^6 
  \left(\frac{\epsilon_{\rm X}}{1\mbox{ keV}}\right)^{1/2}
  \left(\frac{B_2}{5.5\mbox{ G}}\right)^{-1} \quad \mbox{cm},\\
\end{equation}
where $\epsilon_{\rm X}$ is the synchrotron photon energy.  The obtained
gyration radius is far smaller than the companion, $r_{\rm comp} \sim
10^{10}$ cm, therefore the shock heated electrons cannot escape from the
downstream immediately.  In terms of the enegetics, the phase averaged
X-ray luminosity of the power-law component was $L_{\rm pow,0.5-10keV}
=3.6\times 10^{31}$ erg s$^{-1}$ from the spectral fitting.  Taking into
account the Lorentz boost effect of 1/0.42, and the ratio of the maximum
flux against the phase averaged flux of $\sim 3.0$, the intrinsic
synchrotron luminosity from the companion becomes $L_{\rm
pow,0.5-10keV}' = 1.1 \times 10^{32}$ which is comparable to the
bolometric luminosity of the heated hemisphere of the companion, $L_{\rm
BB} \sim 1.1\times10^{32}$ ergs s$^{-1}$.  This brief estimation also
indicates that the heating efficiency should be smaller than 0.5.  This
result obviously different from the heating efficiency obtained from the
phase-resolved SED (see Figure \ref{3D_SED_fit}) in which an isotropic
pulsar wind is assumed.  This discrepancy may indicate that the pulsar
wind from the central MSP is not isotropic as expected in the crab
nebula\citep{2002AstL...28..373B}.  The wind pressure at the companion
is then $p_{\rm wind} = L_{\rm SD}/4\pi R_{\rm orb}^2 c \sim 4.7$ dyn
which may be greater than that of the coronal gas ($T=10^6$ K, $n_{\rm
p}=10^8$ cm$^{-3}$ for the sun) but much smaller than the photosphere.
Therefore the thickness of the emitting region should be not higher than
the corona of the companion.  On the other hand the radiation length of
the relativistic electrons is quite long:
$716A/(Z(Z+1)\ln(278/\sqrt(Z))\sim 4\times 10^{17}$ cm for the coronal
gas of $n_{\rm p}\sim10^8$cm$^{-3}$.  Therefore these electrons can
radiate synchrotron X-rays even in the immediate area of the
photosphere.


If the hard X-rays originated in the shocked pulsar wind on the surface
of the companion star as is discussed, the composition of the pulsar
wind can be constrained based on the downstream flow velocity.
Considering the Rankin-Hugoniot relation for magnetized
electron-positron plasma for small $\sigma$, we can find the downstream
flow velocity as a function of the magnetization parameter $\sigma$
\citep{1984ApJ...283..694K},
\begin{equation}
 u_{\rm 2}^2 = \frac{1+9\sigma}{8},
\end{equation}
where $u_{\rm 2} = \Gamma_{\rm 2} \beta_{\rm 2}$.  Finally we find the
downstream flow velocity, $\beta_{\rm 2} =
\sqrt{(1+9\sigma)/(7-9\sigma)}$ in units of the light speed.  Note that
the equation predicts the lower limit of the shocked pulsar wind,
$\beta_{\rm 2} = 1/\sqrt{7} \sim 0.38$ at $\sigma = 0$.  On the other
hand, the flow velocity of $\beta_{\rm 2} = 0.43 \pm 0.15$ from the
X-ray light curve indicates that the pulsar wind is already in the
particle dominant state, i.e., $\sigma \sim 0.028$, at the surface of
the companion, only $1.1\times10^{11}$ cm apart from the pulsar.  Note
that the post-shock flow velocity has an inverse-correlation with the
inclination angle that affects the resultant magnetization parameter.
If we adopt an error range of inclination angle, $50\degr \sim 60\degr$,
the magnetization parameter can range from 0.008 (for $i=60\degr$) to
0.1 (for $i=50\degr$).  The upper limit of $\sigma < 0.1$ is still much
smaller than that calculated for the similar system of PSR J1023+0038
reported by \citet{2011ApJ...742...97B}, in which a thick post-shock
emitting region was assumed.  Past theoretical studies claimed that the
pulsar wind is highly magnetized at the beginning just around the light
cylinder.  However, the observed $\sigma$ is far smaller than 1, and
this discrepancy has been called the $\sigma$ paradox.  Recently
\cite{2012Natur.482..507A} reported evidence of particle acceleration
just around the pulsar based on gamma-ray pulse analysis of the Crab
pulsar.  The above argument of $\sigma$ in this work is consistent with
their result and may be one of the $\sigma$ parameters measured at the
nearest distance from a pulsar.  Furthermore this may imply that the old
MSPs and the young radio pulsars have the same particle acceleration
mechanism.

One of the intriguing features of this object is the double-peak
structure in the 2$-$4 keV X-ray light curve.  Since the dip is observed
only in the softer band (Figure \ref{folded-XLC}), this feature can be
interpreted as absorption of soft X-rays.  If we accept that the
non-thermal X-rays originate in the shocked pulsar wind on the surface
of the companion as discussed above, the absorber must be lying around
the pulsar, because the dip appears at the optical maximum phase (= the
superior conjunction).  In addition, the orbital inclination angle $i
\sim 55\degr$ requires that the absorber must be extended far from
the orbital plane ($\sim 8\times 10^{10}$ cm).

As \cite{2014AAS...22314007R} claimed, mass accretion on the pulsar
seems to continue intermittently, however the argument of the propeller
effect rules out a powerful accretion.  This is consistent with the lack
of a bright disc component in X-ray spectra.  In addition to the
propeller effects, the measured companion radius is about 1/2 of the
Roche lobe, therefore the stellar material cannot overcome the Lagrange
point L1 to accrete on the neutron star.  To feed mass to the pulsar,
explosive ejection faster than $\sim 1000$ km s$^{-1}$ may be required
for the mass to escape from the gravity potential.

In addition, as inferred from the discussion on the heating efficiency,
the pulsar wind may be concentrated on the orbital plane, which would
prevent the accretion flow.  If the stellar material that is blown away
by the equatorial pulsar wind is drifting around the pulsar's polar
regions, and it may accrete from the poloidal direction.  On the other
hand, the running periodogram shown in Figure \ref{fig:Lomb} indicates
that the dip structure related to the second harmonic component is
unstable, and appears to have disappeared $\sim 120$ ks from the start
time of the observation.  To confirm the scenario in which the stellar
gas is drifting around the pulsar and accretes intermittently,
simultaneous X-ray and radio observations are required.

\section{Conclusion}
We presented multi-wavelength observations of a newly found black-widow
binary system 2FGL J2339.6-0532 covering near-infrared to X-ray regimes.
Thanks to a wide wave waveband and long coverage, we successfully
obtained a phase-resolved SED that enabled us to constrain the orbital
parameters more precisely.  The obtained SED seems consistent with past
studies, and we calculated an inclination angle of $i \sim 55\degr$,
while taking into account the results of recent radio observations.
Based on the argument of the propeller-effect argument, the lower limit
of the acrretion rate was estimated to be $\dot{m} > 1.3\times 10^{16}$
g s$^{-1}$, which is about five orders of magnitude larger than that
expected from the observed X-ray luminosity.  In addition, the estimated
orbital parameters imply that the companion's radius is only 1/2 of the
Roche lobe, making it difficult to feed the pulsar continuously.  We
also obtained an H$\alpha$ image of the vicinity of the target and could
not detect any diffuse structure with a 3$\sigma$ detection limit of $<
8.7 \times 10^{-17}$ erg s$^{-1}$ cm$^{-2}$ arcsec$^{-2}$.  We therefore
conclude that the target does not have a bow-shock nebula brighter than
the H$\alpha$ nebula around PSR B1957+20.

In the X-ray regime, we discovered a steady, soft X-ray component below
1 keV, which seemed to originate from the neutron-star surface, and
which showed no evidence of an accretion disc.  On the other hand, the
hard component above 2 keV showed periodic modulation synchronized with
the orbital motion, implying that the hard X-rays originate in the
shocked pulsar wind near the companion surface.  The observed X-ray
luminosity is comparable to the bolometric luminosity of the heated
hemisphere of the companion.  This means that the heating efficiency
should be smaller than 0.5 and the pulsar-wind distribution should be
anisotropic.  Adopting the above scenario we estimated the magnetization
parameter of the pulsar wind $\sigma \sim 0.03$, based on the
Rankin$-$Hugoniot relation.  This implies that the pulsar wind is
already in the particle-dominant state at a distance of 1.1
$\times10^{11}$ cm from the pulsar.  Moreover, we also investigated the
time variability of the modulation pattern by using the running
periodogram of X-ray light curve, wherein we detected a weakening of the
modulation pattern.  This may be related to an intermittent weak
accretion or an unstable pulsar-wind activity, both of which can cause
temporal extinction of the radio-pulse emission.



\acknowledgments This research has made use of optical \& near infrared
data obtained from Optical \& Near-Infrared Astronomy Inter-University
Cooperation Program, supported by the MEXT of Japan and Kottamia
Astronomical Observatory supported by the NRIAG of Egypt.  This research
has also made use of X-ray data obtained from the Suzaku satellite, a
collaborative mission between the space agencies of Japan (JAXA) and the
USA (NASA).  Operation of ANIR on the miniTAO telescope is supported by
Grant-in-Aid for Scientific Research (21684006, 22253002, and 22540258)
and the Institutional Program for Young Researcher Overseas Visits,
operated by Japan Society for the Promotion of Science (JSPS).  Part of
this work has been supported by NAOJ Research Grant for Universities.
The author would like to thank the referee for useful suggestions that
helped to improve the original manuscrip.  YY is deeply grateful to
Suguru Saito for the generous support and encouragement.



{\it Facilities:} \facility{Suzaku}, \facility{OISTER},
\facility{Kottamia Astronomical Observatory}.

\mbox{~}
\bibliography{refs}

\begin{thebibliography}{47}
\expandafter\ifx\csname natexlab\endcsname\relax\def\natexlab#1{#1}\fi

\bibitem[{{Abdo} {et~al.}(2010){Abdo}, {Ackermann}, {Ajello}, {Allafort},
  {Antolini}, {Atwood}, {Axelsson}, {Baldini}, {Ballet}, {Barbiellini}, \&
  et~al.}]{2010ApJS..188..405A}
{Abdo}, A.~A., {et~al.} 2010, \apjs, 188, 405

\bibitem[{{Ackermann} {et~al.}(2012){Ackermann}, {Ajello}, {Allafort},
  {Antolini}, {Baldini}, {Ballet}, {Barbiellini}, {Bastieri}, {Bellazzini},
  {Berenji}, {Blandford}, {Bloom}, {Bonamente}, {Borgland}, {Bouvier},
  {Brandt}, {Bregeon}, {Brigida}, {Bruel}, {Buehler}, {Burnett}, {Buson},
  {Caliandro}, {Cameron}, {Caraveo}, {Casandjian}, {Cavazzuti}, {Cecchi}, {{\c
  C}elik}, {Charles}, {Chekhtman}, {Chen}, {Cheung}, {Chiang}, {Ciprini},
  {Claus}, {Cohen-Tanugi}, {Conrad}, {Cutini}, {de Angelis}, {DeCesar}, {De
  Luca}, {de Palma}, {Dermer}, {Silva}, {Drell}, {Drlica-Wagner}, {Dubois},
  {Enoto}, {Favuzzi}, {Fegan}, {Ferrara}, {Focke}, {Fortin}, {Fukazawa},
  {Funk}, {Fusco}, {Gargano}, {Gasparrini}, {Gehrels}, {Germani}, {Giglietto},
  {Giordano}, {Giroletti}, {Glanzman}, {Godfrey}, {Grenier}, {Grondin},
  {Grove}, {Guillemot}, {Guiriec}, {Gustafsson}, {Hadasch}, {Hanabata},
  {Harding}, {Hayashida}, {Hays}, {Healey}, {Hill}, {Horan}, {Hou},
  {J{\'o}hannesson}, {Johnson}, {Johnson}, {Kamae}, {Katagiri}, {Kataoka},
  {Kerr}, {Kn{\"o}dlseder}, {Kuss}, {Lande}, {Latronico}, {Lee},
  {Lemoine-Goumard}, {Longo}, {Loparco}, {Lott}, {Lovellette}, {Lubrano},
  {Madejski}, {Mazziotta}, {McEnery}, {Mehault}, {Michelson}, {Mignani},
  {Mitthumsiri}, {Mizuno}, {Monte}, {Monzani}, {Morselli}, {Moskalenko},
  {Murgia}, {Nakamori}, {Naumann-Godo}, {Nolan}, {Norris}, {Nuss}, {Ohsugi},
  {Okumura}, {Omodei}, {Orlando}, {Ormes}, {Ozaki}, {Paneque}, {Panetta},
  {Parent}, {Pelassa}, {Pesce-Rollins}, {Pierbattista}, {Piron}, {Pivato},
  {Porter}, {Rain{\`o}}, {Rando}, {Ray}, {Razzano}, {Reimer}, {Reimer},
  {Reposeur}, {Romani}, {Sadrozinski}, {Salvetti}, {Saz Parkinson}, {Schalk},
  {Sgr{\`o}}, {Shaw}, {Siskind}, {Smith}, {Spandre}, {Spinelli}, {Suson},
  {Takahashi}, {Tanaka}, {Thayer}, {Thayer}, {Thompson}, {Tibaldo}, {Tibolla},
  {Torres}, {Tosti}, {Tramacere}, {Troja}, {Usher}, {Vandenbroucke},
  {Vasileiou}, {Vianello}, {Vilchez}, {Vitale}, {Waite}, {Wallace}, {Wang},
  {Winer}, {Wolff}, {Wood}, {Wood}, {Yang}, \& {Zimmer}}]{2012ApJ...753...83A}
{Ackermann}, M., {et~al.} 2012, \apj, 753, 83

\bibitem[{{Aharonian} {et~al.}(2012){Aharonian}, {Bogovalov}, \&
  {Khangulyan}}]{2012Natur.482..507A}
{Aharonian}, F.~A., {Bogovalov}, S.~V., \& {Khangulyan}, D. 2012, \nat, 482,
  507

\bibitem[{{Alpar} {et~al.}(1982){Alpar}, {Cheng}, {Ruderman}, \&
  {Shaham}}]{1982Natur.300..728A}
{Alpar}, M.~A., {Cheng}, A.~F., {Ruderman}, M.~A., \& {Shaham}, J. 1982, \nat,
  300, 728

\bibitem[{{Azzam} {et~al.}(2014){Azzam}, {Ali}, {Elnagahy}, {Ismail}, {Haroon},
  {Selim}, \& {Ahmed-Essam}}]{2014arXiv1402.2926A}
{Azzam}, Y.~A., {Ali}, G.~B., {Elnagahy}, F., {Ismail}, H.~A., {Haroon}, A.,
  {Selim}, I., \& {Ahmed-Essam}. 2014, ArXiv e-prints

\bibitem[{{Becker} \& {Truemper}(1997)}]{1997A&A...326..682B}
{Becker}, W., \& {Truemper}, J. 1997, \aap, 326, 682

\bibitem[{{Bogdanov} {et~al.}(2011){Bogdanov}, {Archibald}, {Hessels}, {Kaspi},
  {Lorimer}, {McLaughlin}, {Ransom}, \& {Stairs}}]{2011ApJ...742...97B}
{Bogdanov}, S., {Archibald}, A.~M., {Hessels}, J.~W.~T., {Kaspi}, V.~M.,
  {Lorimer}, D., {McLaughlin}, M.~A., {Ransom}, S.~M., \& {Stairs}, I.~H. 2011,
  \apj, 742, 97

\bibitem[{{Bogovalov} \& {Khangoulyan}(2002)}]{2002AstL...28..373B}
{Bogovalov}, S.~V., \& {Khangoulyan}, D.~V. 2002, Astronomy Letters, 28, 373

\bibitem[{{Cardelli} {et~al.}(1989){Cardelli}, {Clayton}, \&
  {Mathis}}]{1989ApJ...345..245C}
{Cardelli}, J.~A., {Clayton}, G.~C., \& {Mathis}, J.~S. 1989, \apj, 345, 245

\bibitem[{{Chatterjee} \& {Cordes}(2002)}]{2002ApJ...575..407C}
{Chatterjee}, S., \& {Cordes}, J.~M. 2002, \apj, 575, 407

\bibitem[{{Dickey} \& {Lockman}(1990)}]{1990ARA&A..28..215D}
{Dickey}, J.~M., \& {Lockman}, F.~J. 1990, \araa, 28, 215

\bibitem[{{Ferrigno} {et~al.}(2014){Ferrigno}, {Bozzo}, {Papitto}, {Rea},
  {Pavan}, {Campana}, {Wieringa}, {Filipovi{\'c}}, {Falanga}, \&
  {Stella}}]{2014A&A...567A..77F}
{Ferrigno}, C., {et~al.} 2014, \aap, 567, A77

\bibitem[{{Fruchter} {et~al.}(1988){Fruchter}, {Stinebring}, \&
  {Taylor}}]{1988Natur.333..237F}
{Fruchter}, A.~S., {Stinebring}, D.~R., \& {Taylor}, J.~H. 1988, \nat, 333, 237

\bibitem[{{Fukugita} {et~al.}(1996){Fukugita}, {Ichikawa}, {Gunn}, {Doi},
  {Shimasaku}, \& {Schneider}}]{1996AJ....111.1748F}
{Fukugita}, M., {Ichikawa}, T., {Gunn}, J.~E., {Doi}, M., {Shimasaku}, K., \&
  {Schneider}, D.~P. 1996, \aj, 111, 1748

\bibitem[{{Gajjar} {et~al.}(2012){Gajjar}, {Joshi}, \&
  {Kramer}}]{2012MNRAS.424.1197G}
{Gajjar}, V., {Joshi}, B.~C., \& {Kramer}, M. 2012, \mnras, 424, 1197

\bibitem[{{Hester} {et~al.}(2002){Hester}, {Mori}, {Burrows}, {Gallagher},
  {Graham}, {Halverson}, {Kader}, {Michel}, \& {Scowen}}]{2002ApJ...577L..49H}
{Hester}, J.~J., {et~al.} 2002, \apjl, 577, L49

\bibitem[{{Itoh} {et~al.}(2001){Itoh}, {Soyano}, {Tarusawa}, {Aoki}, {Yoshida},
  {Hasegawa}, {Yadomaru}, {Nakada}, \& {Miyazaki}}]{2001PNAOJ...6...41I}
{Itoh}, N., {et~al.} 2001, Publications of the National Astronomical
  Observatory of Japan, 6, 41

\bibitem[{{Kalberla} {et~al.}(2005){Kalberla}, {Burton}, {Hartmann}, {Arnal},
  {Bajaja}, {Morras}, \& {P{\"o}ppel}}]{2005A&A...440..775K}
{Kalberla}, P.~M.~W., {Burton}, W.~B., {Hartmann}, D., {Arnal}, E.~M.,
  {Bajaja}, E., {Morras}, R., \& {P{\"o}ppel}, W.~G.~L. 2005, \aap, 440, 775

\bibitem[{{Kargaltsev} \& {Pavlov}(2008)}]{2008AIPC..983..171K}
{Kargaltsev}, O., \& {Pavlov}, G.~G. 2008, in American Institute of Physics
  Conference Series, Vol. 983, 40 Years of Pulsars: Millisecond Pulsars,
  Magnetars and More, ed. {C.~Bassa, Z.~Wang, A.~Cumming, \& V.~M.~Kaspi},
  171--185

\bibitem[{{Kataoka} {et~al.}(2012){Kataoka}, {Yatsu}, {Kawai}, {Urata},
  {Cheung}, {Takahashi}, {Maeda}, {Totani}, {Makiya}, {Hanayama}, {Miyaji}, \&
  {Tsai}}]{2012ApJ...757..176K}
{Kataoka}, J., {et~al.} 2012, \apj, 757, 176

\bibitem[{{Kawabata} {et~al.}(2008){Kawabata}, {Nagae}, {Chiyonobu}, {Tanaka},
  {Nakaya}, {Suzuki}, {Kamata}, {Miyazaki}, {Hiragi}, {Miyamoto}, {Yamanaka},
  {Arai}, {Yamashita}, {Uemura}, {Ohsugi}, {Isogai}, {Ishitobi}, \&
  {Sato}}]{2008SPIE.7014E..4LK}
{Kawabata}, K.~S., {et~al.} 2008, in Society of Photo-Optical Instrumentation
  Engineers (SPIE) Conference Series, Vol. 7014, Society of Photo-Optical
  Instrumentation Engineers (SPIE) Conference Series, 70144L

\bibitem[{{Kennel} \& {Coroniti}(1984)}]{1984ApJ...283..694K}
{Kennel}, C.~F., \& {Coroniti}, F.~V. 1984, \apj, 283, 694

\bibitem[{{Kong} {et~al.}(2012){Kong}, {Huang}, {Cheng}, {Takata}, {Yatsu},
  {Cheung}, {Donato}, {Lin}, {Kataoka}, {Takahashi}, {Maeda}, {Hui}, \&
  {Tam}}]{2012ApJ...747L...3K}
{Kong}, A.~K.~H., {et~al.} 2012, \apjl, 747, L3

\bibitem[{{Konishi}(2014)}]{2014PASJ...accepted}
{Konishi}, M. e.~a. 2014, accepted for PASJ

\bibitem[{{Kotani} {et~al.}(2005){Kotani}, {Kawai}, {Yanagisawa}, {Watanabe},
  {Arimoto}, {Fukushima}, {Hattori}, {Inata}, {Izumiura}, {Kataoka}, {Koyano},
  {Kubota}, {Kuroda}, {Mori}, {Nagayama}, {Ohta}, {Okada}, {Okita}, {Sato},
  {Serino}, {Shimizu}, {Shimokawabe}, {Suzuki}, {Toda}, {Ushiyama}, {Yatsu},
  {Yoshida}, \& {Yoshida}}]{2005NCimC..28..755K}
{Kotani}, T., {et~al.} 2005, Nuovo Cimento C Geophysics Space Physics C, 28,
  755

\bibitem[{{Kulkarni} \& {Hester}(1988)}]{1988Natur.335..801K}
{Kulkarni}, S.~R., \& {Hester}, J.~J. 1988, \nat, 335, 801

\bibitem[{{Landolt}(1992)}]{1992AJ....104..340L}
{Landolt}, A.~U. 1992, \aj, 104, 340

\bibitem[{{Linares} {et~al.}(2014){Linares}, {Bahramian}, {Heinke}, {Wijnands},
  {Patruno}, {Altamirano}, {Homan}, {Bogdanov}, \&
  {Pooley}}]{2014MNRAS.438..251L}
{Linares}, M., {et~al.} 2014, \mnras, 438, 251

\bibitem[{{Longair}(1994)}]{1994hea..book.....L}
{Longair}, M.~S. 1994, {High energy astrophysics. Vol.2: Stars, the galaxy and
  the interstellar medium} (Cambridge: Cambridge University Press, |c1994, 2nd
  ed.)

\bibitem[{{Motohara} {et~al.}(2010){Motohara}, {Konishi}, {Toshikawa},
  {Mitani}, {Sako}, {Uchimoto}, {Yamamuro}, {Minezaki}, {Tanabe}, {Miyata},
  {Koshida}, {Kato}, {Ohsawa}, {Nakamura}, {Asano}, {Yoshii}, {Doi}, {Kohno},
  {Tanaka}, {Kawara}, {Handa}, {Aoki}, {Soyano}, {Tarusawa}, \&
  {Ita}}]{2010SPIE.7735E..3KM}
{Motohara}, K., {et~al.} 2010, in Society of Photo-Optical Instrumentation
  Engineers (SPIE) Conference Series, Vol. 7735, Society of Photo-Optical
  Instrumentation Engineers (SPIE) Conference Series, 77353

\bibitem[{{Nagayama} {et~al.}(2003){Nagayama}, {Nagashima}, {Nakajima},
  {Nagata}, {Sato}, {Nakaya}, {Yamamuro}, {Sugitani}, \&
  {Tamura}}]{2003SPIE.4841..459N}
{Nagayama}, T., {et~al.} 2003, in Society of Photo-Optical Instrumentation
  Engineers (SPIE) Conference Series, Vol. 4841, Instrument Design and
  Performance for Optical/Infrared Ground-based Telescopes, ed. M.~{Iye} \&
  A.~F.~M. {Moorwood}, 459--464

\bibitem[{{Papitto} {et~al.}(2013){Papitto}, {Ferrigno}, {Bozzo}, {Rea},
  {Pavan}, {Burderi}, {Burgay}, {Campana}, {di Salvo}, {Falanga},
  {Filipovi{\'c}}, {Freire}, {Hessels}, {Possenti}, {Ransom}, {Riggio},
  {Romano}, {Sarkissian}, {Stairs}, {Stella}, {Torres}, {Wieringa}, \&
  {Wong}}]{2013Natur.501..517P}
{Papitto}, A., {et~al.} 2013, \nat, 501, 517

\bibitem[{{Pelling} {et~al.}(1987){Pelling}, {Paciesas}, {Peterson},
  {Makishima}, {Oda}, {Ogawara}, \& {Miyamoto}}]{1987ApJ...319..416P}
{Pelling}, R.~M., {Paciesas}, W.~S., {Peterson}, L.~E., {Makishima}, K., {Oda},
  M., {Ogawara}, Y., \& {Miyamoto}, S. 1987, \apj, 319, 416

\bibitem[{{Ray} {et~al.}(2014){Ray}, {Belfiore}, {Saz Parkinson}, {Polisensky},
  {Ransom}, {Romani}, {Hessels}, {Razzano}, {Bhattacharyya}, {Roy}, {Cognard},
  \& {Pulsar Search Consortium}}]{2014AAS...22314007R}
{Ray}, P.~S., {et~al.} 2014, in American Astronomical Society Meeting
  Abstracts, Vol. 223, American Astronomical Society Meeting Abstracts,
  \#140.07

\bibitem[{{Roberts}(2011)}]{2011AIPC.1357..127R}
{Roberts}, M.~S.~E. 2011, in American Institute of Physics Conference Series,
  Vol. 1357, American Institute of Physics Conference Series, ed. M.~{Burgay},
  N.~{D'Amico}, P.~{Esposito}, A.~{Pellizzoni}, \& A.~{Possenti}, 127--130

\bibitem[{{Roberts}(2013)}]{2013IAUS..291..127R}
{Roberts}, M.~S.~E. 2013, in IAU Symposium, Vol. 291, IAU Symposium, 127--132

\bibitem[{{Romani} \& {Shaw}(2011)}]{2011ApJ...743L..26R}
{Romani}, R.~W., \& {Shaw}, M.~S. 2011, \apjl, 743, L26

\bibitem[{{Schlafly} \& {Finkbeiner}(2011)}]{2011ApJ...737..103S}
{Schlafly}, E.~F., \& {Finkbeiner}, D.~P. 2011, \apj, 737, 103

\bibitem[{{Skrutskie} {et~al.}(2006){Skrutskie}, {Cutri}, {Stiening},
  {Weinberg}, {Schneider}, {Carpenter}, {Beichman}, {Capps}, {Chester},
  {Elias}, {Huchra}, {Liebert}, {Lonsdale}, {Monet}, {Price}, {Seitzer},
  {Jarrett}, {Kirkpatrick}, {Gizis}, {Howard}, {Evans}, {Fowler}, {Fullmer},
  {Hurt}, {Light}, {Kopan}, {Marsh}, {McCallon}, {Tam}, {Van Dyk}, \&
  {Wheelock}}]{2006AJ....131.1163S}
{Skrutskie}, M.~F., {et~al.} 2006, \aj, 131, 1163

\bibitem[{{Smith} {et~al.}(2002){Smith}, {Tucker}, {Kent}, {Richmond},
  {Fukugita}, {Ichikawa}, {Ichikawa}, {Jorgensen}, {Uomoto}, {Gunn}, {Hamabe},
  {Watanabe}, {Tolea}, {Henden}, {Annis}, {Pier}, {McKay}, {Brinkmann}, {Chen},
  {Holtzman}, {Shimasaku}, \& {York}}]{2002AJ....123.2121S}
{Smith}, J.~A., {et~al.} 2002, \aj, 123, 2121

\bibitem[{{Stella} {et~al.}(1986){Stella}, {White}, \&
  {Rosner}}]{1986ApJ...308..669S}
{Stella}, L., {White}, N.~E., \& {Rosner}, R. 1986, \apj, 308, 669

\bibitem[{{Takahashi} {et~al.}(2009){Takahashi}, {Nishihara}, \& 150cm
  Telescope Working~Group}]{2009GAO-ITB...T}
{Takahashi}, N., {Nishihara}, E., \& 150cm Telescope Working~Group, G. 2009, in
  Proc. of GAO-ITB Joint Workshop in Astronomy and Science Education, GAO-ITB
  Joint Workshop in Astronomy and Science Education, 50--60

\bibitem[{{Tokunaga} \& {Vacca}(2005)}]{2005PASP..117..421T}
{Tokunaga}, A.~T., \& {Vacca}, W.~D. 2005, \pasp, 117, 421

\bibitem[{{Watanabe} {et~al.}(2012){Watanabe}, {Takahashi}, {Sato}, {Watanabe},
  {Fukuhara}, {Hamamoto}, \& {Ozaki}}]{2012SPIE.8446E..2OW}
{Watanabe}, M., {Takahashi}, Y., {Sato}, M., {Watanabe}, S., {Fukuhara}, T.,
  {Hamamoto}, K., \& {Ozaki}, A. 2012, in Society of Photo-Optical
  Instrumentation Engineers (SPIE) Conference Series, Vol. 8446, Society of
  Photo-Optical Instrumentation Engineers (SPIE) Conference Series

\bibitem[{{Weisskopf} {et~al.}(2000){Weisskopf}, {Hester}, {Tennant}, {Elsner},
  {Schulz}, {Marshall}, {Karovska}, {Nichols}, {Swartz}, {Kolodziejczak}, \&
  {O'Dell}}]{2000ApJ...536L..81W}
{Weisskopf}, M.~C., {et~al.} 2000, \apjl, 536, L81

\bibitem[{{Yanagisawa} {et~al.}(2010){Yanagisawa}, {Kuroda}, {Yoshida},
  {Shimizu}, {Nagayama}, {Toda}, {Ohta}, \& {Kawai}}]{2010AIPC.1279..466Y}
{Yanagisawa}, K., {Kuroda}, D., {Yoshida}, M., {Shimizu}, Y., {Nagayama}, S.,
  {Toda}, H., {Ohta}, K., \& {Kawai}, N. 2010, in American Institute of Physics
  Conference Series, Vol. 1279, American Institute of Physics Conference
  Series, ed. N.~{Kawai} \& S.~{Nagataki}, 466--468

\bibitem[{{Yoshida}(2005)}]{2005JKAS...38..117Y}
{Yoshida}, M. 2005, Journal of Korean Astronomical Society, 38, 117

\end{thebibliography}
\mbox{~}
\tabletypesize{\tiny}
\begin{deluxetable*}{lc|ccccccc|cccc}
\tablecolumns{13}
\tablecaption{Summary of optical observations with OISTER
\label{OISTER_obs_summary}}
\tablewidth{0pt}
\tablehead{
 \colhead{Telescope (Instrument)} & 
 \colhead{Diameter} & 
 \multicolumn{7}{c}{September 2011} &
 \multicolumn{4}{c}{October 2011}\\
 \cline{3-9} \cline{10-13}\\
 \colhead{}&
 \colhead{}&
 \colhead{22}&
 \colhead{23} & 
 \colhead{24} & 
 \colhead{27} &
 \colhead{28} &
 \colhead{29} &
 \colhead{30} &
 \colhead{1} &
 \colhead{4} &
 \colhead{6} &
 \colhead{7} 
}
\startdata
 Pirka telescope (MSI)\tablenotemark{a}&
 160cm&
 & 
 BVRI\tablenotemark{$\dagger$}&
 BVRI\tablenotemark{$\dagger$}& 
 V&
 V&
 &
 &
 &
 &
 &
 \\
 Gunma Astronomical Observatory (GIRCS)\tablenotemark{b} & 
 150cm&
 &
 &
 &
 &
 J&
 &
 &
 &
 &
 &
 \\
 MITSuME-Akeno telescope (Tricolor Camera)\tablenotemark{c} &
 50cm&
 gRI&
 gRI&
 gRI&
 gRI&
 gRI&
 gRI&
 gRI&
 &
 &
 &
 \\
 Kiso Schmidt (2kCCD)\tablenotemark{d} & 
 105cm&
 &
 &
 &
 &
 &
 R\/H$\alpha$ &
 &
 &
 &
 &
 \\
 Kyoto Sangyo University 1.3m telescope (CCD)\tablenotemark{e} & 
 130cm&
 &
 &
 &
 &
 gz&
 &
 &
 &
 &
 &
 \\
 Nayuta telescope (NIC)\tablenotemark{f} & 
 200cm&
 (JHK$_{\rm S}$)&
 &
 &
 &
 &
 &
 &
 &
 &
 &
 \\
 MITSuME-OAO telescope(Tricolor Camera)\tablenotemark{c} & 
 50cm &
 &
 gRI &
 &
 gRI &
 gRI &
 &
 &
 gRI &
 &
 gRI &
 gRI \\
 OAO 188cm telescope (ISLE)\tablenotemark{g} & 
 188cm&
 &
 &
 &
 J&
 &
 &
 &
 &
 &
 &
 \\
 Bisei Spaceguard Center 1m telescope (Optical CCD)\tablenotemark{h} & 
 100cm&
 &
 &
 &
 (r)&
 (r)&
 &
 &
 &
 &
 &
 \\
 Kanata telescope (HOWPol)\tablenotemark{j} & 
 150cm&
 &
 &
 &
 R&
 R&
 &
 &
 &
 V&
 &
 B\\
 Kagoshima University 1m telescope (IR CCD)\tablenotemark{j} & 
 100cm&
 &
 &
 &
 (J)&
 (J)&
 &
 &
 (J)&
 &
 &
 \\
 Murikabushi telescope (Tricolor Camera)\tablenotemark{c} & 
 105cm&
 &
 &
 &
 gRI&
 &
 gRI&
 gRI&
 &
 &
 &
 \\
 Kottamia Astronomical Observatory (Optical CCD)\tablenotemark{k} & 
 188cm&
 &
 &
 &
 R&
 R&
 R&
 B&
 &
 &
 &
 \\
 IRSF 1.4m telescope (SIRIUS)\tablenotemark{l} & 
 140cm&
 &
 &
 &
 &
 JHK$_{\rm S}$ &
 JHK$_{\rm S}$ &
 JHK$_{\rm S}$ &
 &
 &
 &
 \\
 miniTAO (ANIR)\tablenotemark{l}& 
 104cm&
 &
 &
 &
 &
 &
 RJ&
 BJ&
 &
 &
 &
 \\
 \enddata
 \tablecomments{The letters on the table describe the observed energy
 bands.  Lower case and upper case characters correspond to the SDSS
 system and the Johnson$-$Cousins system, respectively.  The
 observations shown in parentheses did not produce usable data due to
 the weather condition.}

 \tablenotetext{$\dagger$}{Calibration observation for the field photometry.}
 \tablenotetext{a}{Nayoro, Hokkaido Pref., Japan
 \citep{2012SPIE.8446E..2OW}}
 \tablenotetext{b}{Takayama, Gunma Pref., Japan \citep{2009GAO-ITB...T}}
 \tablenotetext{c}{Hokuto, Yamanashi Pref., Japan
 \citep{2005NCimC..28..755K}} 
 \tablenotetext{d}{Kiso, Nagano Pref., Japan \citep{2001PNAOJ...6...41I}}
 \tablenotetext{e}{Kyoyo, Kyoto Pref., Japan
 (\url{http://www.kyoto-su.ac.jp/kao/})} 
 \tablenotetext{f}{Sayo, Hyogo Pref., Japan
 (\url{http://www.nhao.jp/en/})} 
 \tablenotetext{g}{Asaguchi, Okayama Pref., Japan
 \citep{2010AIPC.1279..466Y}} 
 \tablenotetext{h}{Asaguchi, Okayama Pref., Japan
 \citep{2005JKAS...38..117Y}} \tablenotetext{i}{Bisei, Okayama Pref.,
 Japan (\url{http://www.spaceguard.or.jp/bsgc\_jsf/pamphlet/index.htm})}
 \tablenotetext{j}{Higashi-hiroshima, Hiroshima Pref., Japan
 \citep{2008SPIE.7014E..4LK}} 
 \tablenotetext{k}{Satsuma-sendai, Kagoshima Pref., Japan
 (\url{http://milkyway.sci.kagoshima-u.ac.jp/1m/sys/index.html})}
 \tablenotetext{c}{Ishigaki, Okinawa Pref., Japan
 \citep{2005NCimC..28..755K}} 
 \tablenotetext{l}{Kottamia, Egypt \citep{2014arXiv1402.2926A}}
 \tablenotetext{m}{Sutherland, South Africa \citep{2003SPIE.4841..459N}}
 \tablenotetext{n}{Atacama, Chile
 \citep{2014PASJ...accepted,2010SPIE.7735E..3KM}}
\end{deluxetable*}

\end{document}